\documentclass[11pt]{article}   	
\usepackage[utf8]{inputenc}
\usepackage{setspace}

\usepackage[top=1in,bottom=1in]{geometry}                		
\usepackage{graphicx}				
		
\usepackage{booktabs} 
\usepackage{array} 
\usepackage{paralist} 
\usepackage{verbatim} 
\usepackage{subfig} 
\usepackage{amsmath}
\usepackage{amssymb}
\usepackage{amsthm}
  {
      \theoremstyle{plain}
      
  }
\usepackage{tikz}
\usepackage{tikz-cd}
\usetikzlibrary{patterns}
\usetikzlibrary{calc}
\usepackage{eulervm}
\usetikzlibrary{calc}
\usepackage{blkarray}
\usepackage{float}
\usepackage{mdframed}
\usepackage{framed}
\usepackage{rotating}
\usepackage{subfig}
\usepackage{physics}

\usetikzlibrary{decorations.pathreplacing,angles,quotes}

\usepackage{mathpazo}
\newcommand{\supp}{\operatorname{supp}} 
\DeclareSymbolFont{Symbols}{OMS}{zplm}{m}{n}
\DeclareMathSymbol{\infty}{\mathord}{Symbols}{"31}

\usepackage{fancyhdr} 
\newtheorem{defi}{Definition}
\newtheorem{theorem}{Theorem}[section]

\newtheorem{proposition}{Proposition}

\newtheorem{observation}{Observation}

\usepackage{lipsum}

\usepackage{hyperref}
\hypersetup{
  colorlinks   = true,    
  urlcolor     = blue,    
  linkcolor    = blue,    
  citecolor    = blue      
}

\begin{document}

\title{Mediated Persuasion}
\author{Andrew Kosenko\footnote{kosenko.andrew@gmail.com; I am deeply grateful to Navin Kartik, for his invaluable help and advice. I would also like to thank Yeon-Koo Che and Joseph Stiglitz for guidance and comments from which I have benefited immensely, as well as James Best, Ambuj Dewan, Laura Doval, Nate Neligh, Anh Nguyen, Pietro Ortoleva, Daniel Rappoport, Luca Rigotti, Teck Yong Tan, Roee Teper, Richard van Weelden, and the participants of the Columbia Microeconomic Theory Colloquium and the Pitt Micro Theory Brown Bag for discussions and input. The usual disclaimer applies.}\\\textit{Department of Economics, Accounting, and Finance, Marist College}}

\date{This draft: \today\\First draft: October 2017}	

\maketitle

\begin{abstract}
We study a game of strategic information design between a sender, who chooses state-dependent information structures, a mediator who can then garble the signals generated from these structures, and a receiver who takes an action after observing the signal generated by the first two players. We characterize sufficient conditions for information revelation, compare outcomes with and without a mediator and provide comparative statics with regard to the preferences of the sender and the mediator. We also provide novel conceptual and computational insights about the set of feasible posterior beliefs that the sender can induce, and use these results to obtain insights about equilibrium outcomes. The sender never benefits from mediation, while the receiver might. The receiver benefits when the mediator's preferences are \textit{not} perfectly aligned with hers; rather the mediator should prefer more information revelation than the sender, but less than perfect revelation.
\end{abstract}

\vspace{1.5in}

\textbf{JEL Classification}: D82, D83, C72.

\textbf{Keywords}: persuasion, strategic communication, information transmission, intermediation, noisy communication, Blackwell informativeness, information garbling, strategic information provision. 

\pagebreak

\begin{section}{Introduction and Literature Setting}

How does the presence of a mediator affect the informational interaction between two parties? In this paper we study a game of persuasion between one side (a sender) that is trying to persuade another side (a receiver) to take a certain action; we add to this standard environment a mediator who is able to alter the recommendation of the sender in some way, before the receiver takes her action. 

The paper has two main contributions. One is technical and concerns computing the set of receiver posterior beliefs that can be induced; we introduce a novel way of solving this problem and one that can be used in other settings. In addition, there is a strong parallel between our results, their representation and implications, and the literature on ordering information structures.\footnote{We pursue this line of inquiry in a contemporaneous related paper.} The second, more important, contribution concerns equilibria of the mediated persuasion game. We consider several economically important classes of utilities (namely, linear, concave, convex and step functions) and provide results about information revelation and welfare in equilibrium for those classes of utilities. Notably, while most papers in this literature focus on the sender's most preferred equilibria, taking the view that the sender can "steer" the receiver into the appropriate equilibrium, we also seriously consider the welfare of the receiver across different equilibria, taking the view that in most applications, it is the \textit{receiver's} welfare that one cares about ultimately. 

The subject of persuasion, broadly construed, is currently being actively investigated in information economics; much excellent research has been produced in the last few years on this, and the topic is continuing to prove a fertile ground for models and applications. More particularly, the topic of information design - the study of how information \textit{endogenously} affects incentives and vice versa - is swiftly becoming a major avenue of research. We add an \textit{institutional} aspect to this research program, and investigate the effects of different informational-organizational topologies on information revelation and welfare. 

In the model studied here, the sender and the receiver are restricted to communicate indirectly, via an intermediator (perhaps more than one), due to technical or institutional constraints. For example, when a financial firm issues certain kinds of financial products, some large (institutional) investors are prohibited from purchasing them, unless they have been rated by a third party, and have achieved a certain rating. Similarly, in many organizations (including many firms, the military, and the intelligence community) the flow of information is directed, with the direction exogenously predetermined, with various agents having the ability to alter (or perhaps not pass on) the information passed up to them. This is precisely the kind of setting we are concerned with here. 

The problem as we have formulated it is quite difficult in general. Here, one strategic player (the sender) can both create and destroy information, and the other strategic player (the mediator) can only destroy information that the first player provided, but cannot create any new information, and in addition, both\footnote{There is of course the third player - the receiver - but she is effectively facing a decision problem. The key interplay is between the sender and the mediator.} of these players only have probabilistic control over evidence realization. 

One of the major difficulties is that when the mediator changes her action, not only does the sender's best response generically change (this is, of course, at the heart of all game-theoretic models), but the effective choice set of the sender changes as well. We overcome this problem and show how to compute explicitly the feasible sets for the mediator the and the sender as functions of the sender's and the mediator's actions.

As a consequence of the modeling assumptions, it will turn out that given the choice of the sender,  the mediator can deviate to anything less informative in the sense of David Blackwell, while given the choice of the mediator the sender can deviate to \textit{some} things that are less Blackwell informative (than the implied final experiment), but not everything. What these "some" things are is precisely what we characterize in the first part of the paper. 

This is also what makes our work different. The major thrust of the literature on multi-sender Bayesian persuasion has focused on players only being able to add (in a certain sense) information.\footnote{For example, in the Gentzkow and Kamenica work on this problem they identify a condition - Blackwell-connectedness - which ensures that full revelation is always an equilibrium outcome in their multisender game. The condition says that given the information provided by the others, any individual player can always deviate to something \textit{more} informative. Li and Normal similarly assume that each sequential player has access to signals that are arbitrarily correlated, so that a player can improve upon the information provided by others realization by realization. In both cases any one player can unilaterally increase the amount of information provided.} On the other hand, the contemporaneous work on persuasion with noise (where some information is exogenously destroyed) has studied nonstrategic settings. We consider an environment where some players can add information, some can subtract information, and in addition, we study a game, not a decision problem. Furthermore, we compare outcomes of the game along two dimensions; first we vary the preferences of the sender and he mediator. The most prominent result is that (perhaps unsurprisingly) preference divergence "quickly" leads to the only equilibrium being uninformative. The final object of this exercise is to compare outcomes (for fixed preferences), in terms of information revelation and welfare, between standard Bayesian persuasion, and Bayesian persuasion with an informational mediator. We show that although the mediator can only destroy information (in an appropriate sense), this can still result in a strict increase in the amount of information revealed in a very strong sense - Blackwell dominance. In simple, common, and non-pathological environments we show that mediation cannot lead to an increase in information revelation. Moreover, in these environments mediation of the sort we discuss has unambiguous detrimental effects on the welfare of the key players. This is not, however, true in general, as we show by illuminating examples. 

A "complete" solution to this problem is, of course, the following: explicitly exhibit the actions chosen, and the equilibrium outcomes (belief distributions) as functions of arbitrary utilities of the players. We will not solve the problem at this level of generality. Rather, we will solve some economically important special cases, and comment informally upon features of the general outcomes in the conclusion. 

Our work provides a foundation for analyzing when informational mediation of the sort we discuss is actually beneficial to the receiver.\footnote{Obviously, it is never strictly beneficial to the sender.} In other words, given some preferences for the players, when would the receiver (with those preferences) prefer to play the game with a mediator to playing the same game with the same preferences but without a mediator? The answer, perhaps unsurprisingly, is sometimes yes, sometimes no, depending on the preferences. The second main contribution of the paper is in providing examples and analyzing some important base cases, such as when the relevant utilities are linear, strictly concave and convex, or are step functions. 

This work is at the intersection of two literatures - strategic information design and noisy persuasion/communication. Our work relies on some results, and is in the spirit of, the celebrated "Bayesian persuasion" approach of \hyperlink{Kamenica and Gentzkow (2011)}{Kamenica and Gentzkow (2011)} (referred to simply as "KG" for brevity hereafter) who consider a simpler version of this problem, and discuss an application of a certain concavification result first considered in chapter 1 of \hyperlink{Aumann and Maschler (1995)}{Aumann and Maschler (1995)}. \hyperlink{Sah and Stiglitz (1986)}{Sah and Stiglitz (1986)} introduced the analysis of economic systems organized in parallel and in series; hierarchies and polyarchies of persuasion via provision of information have already been explored in previous work (\hyperlink{Gentzkow and Kamenica (2017a)}{Gentzkow and Kamenica (2017a)} (referred to as "GK" henceforth, not to be confused with "KG"), 

There are a number of papers that are closely related to the present model. One is \hyperlink{Ambrus, Azevedo and Kamada (2013)}{Ambrus, Azevedo and Kamada (2013)} which considers a cheap talk model where the sender and receiver also communicate via chains of intermediators. Our work is similar in that talk is "cheap" here as well, meaning that the specific choices of the sender and the mediators do not enter their utility functions directly and only do so through the action of the receiver; in addition, we, too, have an analogous communication sequence. The difference is that the sender is not perfectly informed about the state, the message she sends depends on the state, and is in general, stochastic. \hyperlink{Li and Norman (2018)}{Li and Norman (2018)}'s paper on sequential persuasion serves as another stepping stone - they have a very similar model of persuasion, except that the senders move sequentially, observing the history of actions of the senders who moved before them (unlike in our model), and can provide arbitrarily correlated experiments. The other relevant work is \hyperlink{Gentzkow and Kamenica (2017)}{Gentzkow and Kamenica (2017)}'s work on competition in persuasion where the senders move simultaneously (like in our model), but all senders are trying to provide information about the state of the world, whereas we study an environment where the mediator is trying to provide information about the realization of the sender's experiment. \hyperlink{Linpnowski, Ravid and Shishkin (2018)}{Lipnowski, Ravid and Shishkin (2018)} also study a related environment where a "weak institution" in their parlance plays the role of a kind of informational mediator, although the setup is considerably different and there is no role for the interplay of preferences which we focus on here. The subject of introducing a mediator to potentially improve outcomes has also been studied in contract theory (see, inter alia, \hyperlink{Pollrich (2017)}{Pollrich (2017)} and \hyperlink{Rahman and Obara (2010)}{Rahman and Obara (2010)}).

\hyperlink{Perez-Richet and Skreta (2018)}{Perez-Richet and Skreta (2018)} present a complementary model that differs in one key respect - the mediator (using our nomenclature) moves first and her choice is observed by the sender before the sender acts. Our focus is on analyzing outcomes of a particular game as one changes preferences for the mediator (and fixing the signal realization spaces in advance), while they focus on equilibria of a game where the preferences of the mediator are always fully aligned with those of the receiver. More specifically, they construct a "test" where the sender/persuader employs a continuum of signal realizations to pass or fail different types of sender. Plainly, the difference between our work and theirs is that we fix the signal realization space and vary the preferences of the players, while they fix the preferences and derive the optimal signal realization space (and signal realization probabilities). Notably, the contrast with \hyperlink{Perez-Richet and Skreta (2018)}{Perez-Richet and Skreta (2018)} immediately shows that it is strictly with loss of generality to restrict the space of signal realizations, as we do in the paper. This assumption, however, greatly simplifies our problem.\footnote{Indeed, if one were to consider a problem of which both this paper and \hyperlink{Perez-Richet and Skreta (2018)}{Perez-Richet and Skreta (2018)} are special cases, one would have a strategic problem with an unrestricted domain of utilities with complicated infinite-dimensional action spaces.}

\hyperlink{Stulovici (2017)}{Strulovici (2017)} in his "Mediated Truth" paper explores a somewhat related environment where a "mediator" - an expert of some sort or a law enforcement officer - has access to information that is "costly to acquire, cheap to manipulate and produced sequentially". He shows that when information is reproducible and not asymptotically scarce (for example, one can perform many scientific experiments) then societies will learn the truth, while when information is limited (such as evidence from a crime) the answer is negative. In our work we consider a one-shot game, but his insight provides an interesting contrast. For example, a repeated version of the game considered here would satisfy the condition for evidence to not be asymptotically scare, however, it is not clear that this is enough to overcome the incentive problem when the mediator can only garble the signals; certainly there will be no learning is the unique equilibrium in our model is uninformative, as can be the case. 

\hyperlink{Le Treust and Tomala (2018)}{Le Treust and Tomala (2018)} study a very similar, but simpler setting. They consider persuasion with an additional constraint - exogenous noise - and show that while the sender generically suffers a loss as a result of the noise, information-theoretic tools show that the sender can do as well as possible, provided she plays the game enough times (i.e. enough independent copies of the same basic problem are available). Their model can be viewed as a (possibly repeated) special case of the model studied in our paper, with a nonstrategic mediator who chooses a garbling structure that results in the exogenous noise structure. 

\hyperlink{Tsakas and Tsakas (2018)}{Tsakas and Tsakas (2018)} also study the problem of Bayesian persuasion subject to exogenous noise. They show that while it is in principle possible for the sender to benefit from noise, they obtain analogous results to ours (that the sender is always worse off with more noise) when comparing similar noise structures. The reason for why in our model the sender is always worse off, and in their model the sender can be better off is that they consider additional noise structures (which they refer to as "partitional" channels), and the sender may be better off when faced with noise structures that are both canonical and partitional. Thus, our work agrees with theirs along the dimension along which the environments are comparable, but we also consider strategic interaction. 

\hyperlink{Ichihashi (2017)}{Ichihashi (2017)} studies a model in which the sender's information may be limited; he focuses on the cases where doing so might benefit the receiver. In our model a similar role is played by the mediator who modifies the information produced by the sender, and can only modify it by garbling (i.e. only decreasing the amount of information). Thus, while \hyperlink{Ichihashi (2017)}{Ichihashi (2017)} limits the sender's information, we limit what the sender can do with that information. 

We study a game where the players move simultaneously (this is just a modeling trick of course - they do not have to actually act at the same time - the reason for this is because typically one party is not aware of the ratings mechanism or the choice of the financial instruments of the other party when committing to an action; it could also be simply because a player is unable to detect deviations in time to adjust their own strategy); the key point is that the mediator does not see the choice of the sender before making her own choice as in some other models. In other words, we assume "double commitment" - commitment to an information structure for the sender and the mediator, along the lines discussed in KG. This feature generates an interesting possibility of having a kind of prisoner's dilemma not in actions, but in information.\footnote{This is also discussed in GK.} The flow of information is path-dependent (as in \hyperlink{Li and Norman (2018)}{Li and Norman (2018)}), yet not quite sequential while action choices for the sender and the mediator are simultaneous. 

Our focus will be on the amount of information revealed in various organizational setups and the effect of competition and preference (mis)alignment on information revelation and outcomes. Although the basic model is quite general, we have in mind one particular application - the design of a ratings agency. A rating assigned to a financial product can be thought of as an expression of likelihood of default or expected economic loss. A firm (in the parlance of the present setting, the sender) chooses strategically what evidence to submit to a rater (here, the mediator). The mediator, perhaps driven by concerns that may not be identical to those of the firm, then rates the evidence submitted by the firm, and issues a recommendation to the client or the public. We analyze the effect on informativeness and welfare of the mediator's presence in this informational-organizational topology. 

There are several features of this real-world example that deserve mention. First note that the issuing firm itself cannot rate its own financial products; it does, however, \textit{design} its products (or at least gets to choose the products that it submits for review at a particular instance). The ratings agency cannot choose the products - it is constrained to rate the package it has been submitted, but it can choose its ratings process and criteria. It also exhibits the criteria according to which it issued its conclusions. Finally, the purchaser of the financial products (the receiver) is often required to only buy products that have been rated by a reputable firm - in other words, there is an \textit{institutional} constraint at work.

To take a specific example consider structured finance products that consisted of various repackagings of individual loans (mortgages were by far the most important component) into so-called structured investment vehicles, or SIVs. The financial firms issued products that consisted of bundles of individual mortgages, along with rules for obtaining streams of payments from those mortgages. These streams were correlated with each other (since two nearby houses were in the same area, the local economic conditions that affected the ability of one lender to repay, also affected the ability of the other lender to repay), as well as with the overall economy. The firms chose the specific mortgages that went into each SIV strategically. The ratings agencies then rated these SIVs; however, one key element in their ratings (and one that was later shown to be partially responsible for the revealed inaccuracy of those ratings) is that the ratings agencies did not provide their ratings based on the correlations of the returns with the overall economy. Rather, their ratings consisted (mostly) of evaluations of correlations of individual financial products in an SIV with each other. The issuer clearly wants to achieve as high a rating as possible\footnote{And in fact, there is evidence in structured finance that the firms did design their products so that the senior tranches would be as large as possible, while still getting the highest possible rating.}, but if the preferences of the mediator are to "collude" with the seller, this essentially means that there may be very little information revelation in equilibrium.


In this example the state of the world is a complete, fully specified joint distribution of returns; an experiment is a mapping from states of the world into a set that specifies only partial information about the correlations (for example, individual correlations).\footnote{The "big three" firms all utilize fairly coarse scales for ratings.} The mediator then designs a signal (a rating procedure) that maps information about individual correlations into a scaled rating. The precise ratings methodologies are proprietary, so it makes sense to assume that the sender does not know the strategy of the mediator. This example, although it is meant to be suggestive, is not completely analogous to the situation we study. We view the model presented in this paper as a normative exercise, descriptive of interesting features of a problem, but not identical to actual ratings process.

In single-issuer bonds, ratings are mute about correlations with other bonds or with the market. In 2007, less than 1\% of corporate issues but 60\% of all structured products were rated AAA. 27 of 30 AAA issues underwritten by Merril Lynch in 2007, were by 2008 rated as speculative ("junk") (See \hyperlink{Coval, Jurek and Stafford (2008)}{Coval, Jurek and Stafford (2008)}). We suggest that a possible explanation for this is that if the mediator is unable to provide new information, and is only able to "garble" or rely on the information provided to it by the issuer, then the equilibria in general will not be very informative (and in fact, as the preferences of the mediator and the sender diverge, the only equilibrium that survives is uninformative). This reasoning suggests a policy proposal - requiring the ratings agencies to perform independent analysis (say, additional "stress tests") on the products they are rating, to increase the informativeness of the rating; another way of increasing information revelation is to ensure that the preferences of the mediator are what they are prescribed to be by this work.

Another (perhaps more closely parallel, but certainly less important) example of this setting might be the design of a spam filter for an email system.  

In what follows we investigate the effect of adding a mediator to a persuasion environment as well as the welfare implications (for all parties) of varying the alignment of preferences of the sender and the mediator. In addition, we consider the effect of adding additional mediators. Finally, we give a novel characterization of the set of feasible beliefs for this game and discuss its several interesting features. We do not give a full characterization of equilibria as a function of preferences (this is a difficult fixed point problem); rather, we give suggestive examples and provide intuition. 

\begin{section}{Environment}
We study a game with $n \geq 3$ players; The first player is called the \textit{sender} and the last player is called the \textit{receiver}. The remaining players are the \textit{mediators}; if there are more than one of them, we also specify the order in which their probabilistic strategies are executed. 

We fix a finite state space, $\Omega$ (consisting of $n_\Omega$ elements) and a finite realization space\footnote{Typically, the realization space is part of the choice of the sender; here we fix this space (while keeping it "rich enough") to isolate the effects of mediated persuasion.} $E$ (consisting of $n_E$ elements), where to avoid unnecessary trivialities, the cardinality of the set of signals is weakly greater than that of the set of states. An \textit{experiment} for the sender is a distribution over the set $E$, for each state of the world: $X: \Omega \rightarrow \Delta(E)$; denote by $\boldsymbol{X}$ the set of available experiments. We assume that $X$ contains both the uninformative experiment (one where the probabilities of all experiment realizations are independent of the state) and the fully revealing experiment (where each state is revealed with probability one).
 To distinguish between the choices of the sender and those of the mediator, we define a \textit{signal} 
 for the mediator to be a function $\Sigma: E \rightarrow \Delta(S)$ where $S$ is the space of signal realizations containing $n_S$ elements; let $\boldsymbol{\Sigma}$ denote the set of available experiments. Put differently, the mediator is choosing distributions of signal realizations conditional on realizations of experiments. All available experiments and signals have the same cost, which we normalize to zero. We also refer to either an experiment, or a signal, or their product, generically as an \textit{information structure}. Since the state space and all realization spaces are finite, we represent information structures as column-stochastic matrices with the $(i,j)$'th entry being the probability of realization $i$ conditional on $j$. Finally, the receiver takes an action from a finite set $A$ (with $n_A$ elements; we assume that $n_A \geq n_S = n_E \geq n_\Omega$ to avoid trivialities associated with signal and action spaces not being "rich" enough). The utility of the sender is denoted by $\tilde{u}^S(\omega, a)$, that of the mediator by $\tilde{u}^M(\omega, a)$ and that of the receiver by $\tilde{u}^R(\omega, a)$. We assume for concreteness that if the receiver is indifferent between two or more actions given some belief, he takes the action that is best for the sender. 

This setup is capturing one of the key features of our model - the space of realizations of experiments for one player is the state space for the other player. In other words, both the sender and the mediator are choosing standard Blackwell experiments, but with different state and realization spaces. 
 
For clarity, we summarize the notation used at this point: we use the convention that capital Greek letters ($X,\Sigma$) refer to the distributions, bold capital Greek letters ($\boldsymbol{X},\boldsymbol{\Sigma}$) refer to sets of distributions, capital English letters ($E,S$) refer to spaces of realizations for information structures, and small English letters ($e,s$) refer to particular realizations. 

The timing of the game is fairly simple: the sender and the mediator choose their actions simultaneously, while the receiver observes the choices of the experiment, the signal, and the signal realization, but not the experiment realization. The mediator does not observe the choice of the sender when choosing her own action; if he did observe the choice (but not the experiment realization), this would be a special case of the model of sequential persuasion of \hyperlink{Li and Norman (2018)}{Li and Norman (2018)}. If the mediator in addition could observe the experiment realization (and could therefore condition her own action upon it), this would be similar to the models of persuasion with private information by \hyperlink{Hedlund (2017)}{Hedlund (2017)} and \hyperlink{Kosenko (2018)}{Kosenko (2018)}  since then the mediator would have an informational "type". Note that no player observes the realization of the experiment, yet that realization clearly still plays a role in determining outcomes. We focus on pure strategies for all players in the present work; a diagram of the main features, nomenclature, timing, and notational conventions of the model is in figure 1.

\begin{figure}
\begin{tikzpicture}[scale=1.4]
\draw[fill=blue] (3,0) circle [radius=0.1];
\draw [->] (3,0)--(5.8,0);
\draw[fill=blue] (6,0) circle [radius=0.1];
\draw [->] (6,0)--(8.8,0);
\draw[fill=blue] (9,0) circle [radius=0.1];
\draw [->] (9,0)--(11.8,0);
\draw[fill=blue] (12,0) circle [radius=0.1];
\node [above] at (2.8,0.7) {Nature};
\node [below] at (3,0) {State, $\omega$};
\node [below] at (6,-0.2) {$e$};
\node [below] at (9,-0.2) {$s$};
\node [below] at (12,-0.2) {$a$};
\node [above] at (4.5,1) {Sender};
\node [above] at (7.5,1) {Mediator};
\node [above] at (10.5,1) {Receiver};
\node [below] at (4.5,0) {$X$};
\node [below] at (4.5,-0.5) {Experiment};
\node [below] at (7.5,0) {$\Sigma$};
\node [below] at (7.5,-0.5) {Signal};
\end{tikzpicture}
\caption{Illustration of the Model: Flow of Information and Actions.}
\end{figure}

\begin{figure}
\begin{tikzpicture}[scale=1.4]
\draw (0,5)--(10,5);
\draw (0,4.8)--(0,5.2);
\draw (10,4.8)--(10,5.2);
\draw (5,5.2)--(5,4.8);
\node [below] at (5,4.8) {$\beta_0$};
\node [below] at (10,4.7) {1};
\node [below] at (0,4.7) {0};
\draw (0,3)--(10,3);
\draw (2,3.2)--(2,2.8);
\draw (0,2.8)--(0,3.2);
\draw (10,2.8)--(10,3.2);
\node [below] at (2,2.8) {$\beta(X, e_L)$};
\draw (7,3.2)--(7,2.8);
\node [below] at (7,2.8) {$\beta(X, e_H)$};
\node [below] at (0,2.7) {0};
\node [below] at (10,2.7) {1};
\draw (0,1)--(10,1);
\draw (0,0.8)--(0,1.2);
\draw (10,0.8)--(10,1.2);
\node [below] at (0,0.7) {0};
\node [below] at (10,0.7) {1};
\draw (3,1.2)--(3,0.8);
\node [below] at (3,0.7) {$\beta(\Sigma X, s_L)$};
\draw (6,1.2)--(6,0.8);
\node [below] at (6,0.7) {$\beta(\Sigma X, s_H)$};
\draw [dotted] (2,3)--(2,0.8);
\draw [dotted] (7,3)--(7,0.8);
\draw[red,->] (2,0.7)--(3,0.7);
\draw[red,->] (7,0.7)--(6,0.7);
\end{tikzpicture}
\caption{Effect of Garbling on Beliefs in a Dichotomy.}
\end{figure}

The following definition will be extremely useful in what follows:

\begin{defi}
Let $f$ and $g$ be two probability mass functions on a finite set $X=\{x_1,x_2,...,x_k\}\in \mathbb{R}^n$. We say that $f$ is a mean-preserving spread of $g$ if there is a ($k \times k$) Markov matrix $T_{K \times K}=(t(x_i|x_J))_{ij})$ such that 
\begin{enumerate}[i)]
\item
Tg=f.
\item
For each $j=1,...k$, $\sum_iT(x_i|x_j)x_j=x_j$
\end{enumerate}

\end{defi}

We can also illustrate the effect of a garbling of the experiment by the signal on the beliefs (as seen in figure 2). In that figure all players start with a common prior, $\beta_0$. When the sender chooses her experiment $X$, the two possible beliefs (one for each possible realization of the experiment) are a mean-preserving spread of the prior. Following that  mediator's choice of signal, $M$ brings beliefs back in in a mean-preserving contraction. In other words, in terms of figure 2, we can say that the mediator chooses the length (but not the location) of the two arrows, and the sender chooses the outer endpoint for each arrow. The inner point of each arrow represents the final beliefs. 

Denote by $\beta_{A}(\omega|s)$ the posterior belief of the receiver that the state of the world is $\omega$, computed after observing information structure $A$, and a signal realization $s$ and denote by $\beta_{A}(s)$ the full distribution. We will also find it convenient to refer to distributions of distributions, which we will denote by $\tau$ so that $\tau_{A}(\beta)$ is the expected distribution of posterior beliefs given some generic information structure $A$: 
\begin{equation}
\tau_A(\beta)\triangleq \sum_{\{s \in \supp(A)|\beta_A(s)=\beta\} } \hspace{0.3cm} \sum_{\omega \in \Omega}A(s|\omega)\beta_0(\omega)
\end{equation}

We assume that the set of available experiments is anything (or in any case, "rich enough"). In the present work we focus exclusively on pure strategies for all players. This is a major drawback, since as we will see, this environment may have a kind of "matching pennies" flavor where both players constantly want to change their action given what the other is doing (and in particular, finding pure strategy equilibria is quite hard).  Nonetheless we make this restriction for simplicity.

Given a receiver posterior belief (we suppress the arguments for notational compactness) $\beta$, let  $a^*(\beta)$ denote the optimal action of the receiver. Analogously to KG, if two actions for a sender or a mediator result in the same final belief for the receiver, they are equivalent. We can therefore reduce the number of arguments in the utility functions and write $u^R(\beta)$, $u^M(\beta)$, $u^S(\beta)$ (with $u^i(\beta ) \triangleq \mathbb{E}_{\beta} u^i(a^* (\beta),\omega)$, as is customary), and also, with an abuse of notation, $u^R(\tau)$, $u^M(\tau)$, $u^S(\tau)$.




We can begin by observing that an equilibrium exists, and in particular, there is an equilibrium analogous to the "babbling" equilibria of cheap talk models. Suppose for instance, that the sender chooses a completely uninformative experiment. Then the mediator is indifferent between all possible signals, since given the sender's choice, they cannot affect the action of the receiver; in particular he can choose the uninformative signal as well. Clearly, no player can profitably deviate, given the other's choices, and thus this is an equilibrium, which we note in the following

\begin{proposition}
There exists an uninformative equilibrium. 
\end{proposition}

Along the same line of thinking, we have 
\begin{proposition}
Suppose that either $u^S$ or $u^M$ (or both) is globally strictly concave over the set of $\beta \in \Delta (\Omega)$. Then the unique equilibrium is uninformative.
\end{proposition}

The proof of this proposition is immediate from inspection of the utilities (if either utility is concave, then the player with that utility can always bring beliefs back to the prior, which she would prefer to any other outcome); it is also a sufficient condition for the only equilibrium to be uninformative. 

As for nontrivial equilibria, given any $X$, the mediator's problem is now similar to the one faced by the sender in KG: choose a $\Sigma$ such that the distribution of beliefs induced by $B$ is optimal. Formally, the problem for the mediator is:

\begin{equation}
\Sigma^* \in \arg \max_{ \{ \Sigma \in \mathbf{\Sigma}| \Sigma X=B \} } \mathbb{E}_{\tau} u^M(\beta)
\end{equation}
\begin{equation}
\tau=p(B)
\end{equation}

\begin{equation}
\text{ s.t. }\sum_{s \in \supp(\Sigma)}\beta^R(s)\mathbb{P}_B(e)=\beta_0
\end{equation}

Similarly, for the sender the problem is 

\begin{equation}
X^* \in \arg \max_{ \{ X \in \boldsymbol{X}| \Sigma X=B \} } \mathbb{E}_{\tau} u^S(\beta)
\end{equation}
\begin{equation}
\tau=p(B)
\end{equation}
\begin{equation}
\text{ s.t. }\sum_{s \in \supp(\Sigma)}\beta^R(s)\mathbb{P}_B(e)=\beta_0
\end{equation}

Let $p:\mathcal{M}_{n_S,n_{\Omega}} \rightarrow \Delta(\Delta(\Omega))$ where $\mathcal{M}_{n_S,n_\Omega}([0,1])$ denotes the set of $n_S \times n_\Omega$ column-stochastic matrices 
 be the mapping between an information structure and the space of posterior beliefs. In other words, $p$ maps a column-stochastic matrix into a distribution over posteriors: $p(B)=\tau$.

We call a pair $(X,\Sigma)$ that solve the above problems simply an equilibrium and our solution concept is perfect Bayesian equilibrium. We utilize the power of subgame perfection to avoid equilibria in which the receiver threatens to take the worst possible action for the sender unless he observes the fully revealing experiment, and the worst possible action for the mediator unless he observes a fully revealing signal. 

One may notice that the matrix equation $\Sigma X=B$ is precisely the definition for $X$ to be more Blackwell-informative than $B$, with $\Sigma$ being the garbling matrix. We will rely on this fact (as well as the different and related implications of this fact) throughout what is to follow. One can make the simple observation that the set of Blackwell-ranked information structures forms a chain when viewed as a subset of the set of all information structures. 


 


Given a particular choice of $X$ by the sender, the mediator effectively chooses from a set of information structures that are Blackwell-dominated by the experiment. The set of feasible beliefs for the mediator, given a particular choice of the sender is illustrated in figure 3. This set is effectively a proportional "shrinking" of the Bayes-plausible set, since all garblings of $X$ are available to the mediator; the only constraint is that the mediator is not able to induce something more informative (by assumption) than the sender's choice. Jumping ahead we note that the \textit{sender's} feasible set, given a mediator action $\Sigma$ is not going to be a simple "shrinking", and will involve other nontrivial constraints. 

\begin{figure}[h]
\begin{tikzpicture}
 \node[anchor=south west,inner sep=0] at (0,0) {\includegraphics[scale=0.35]{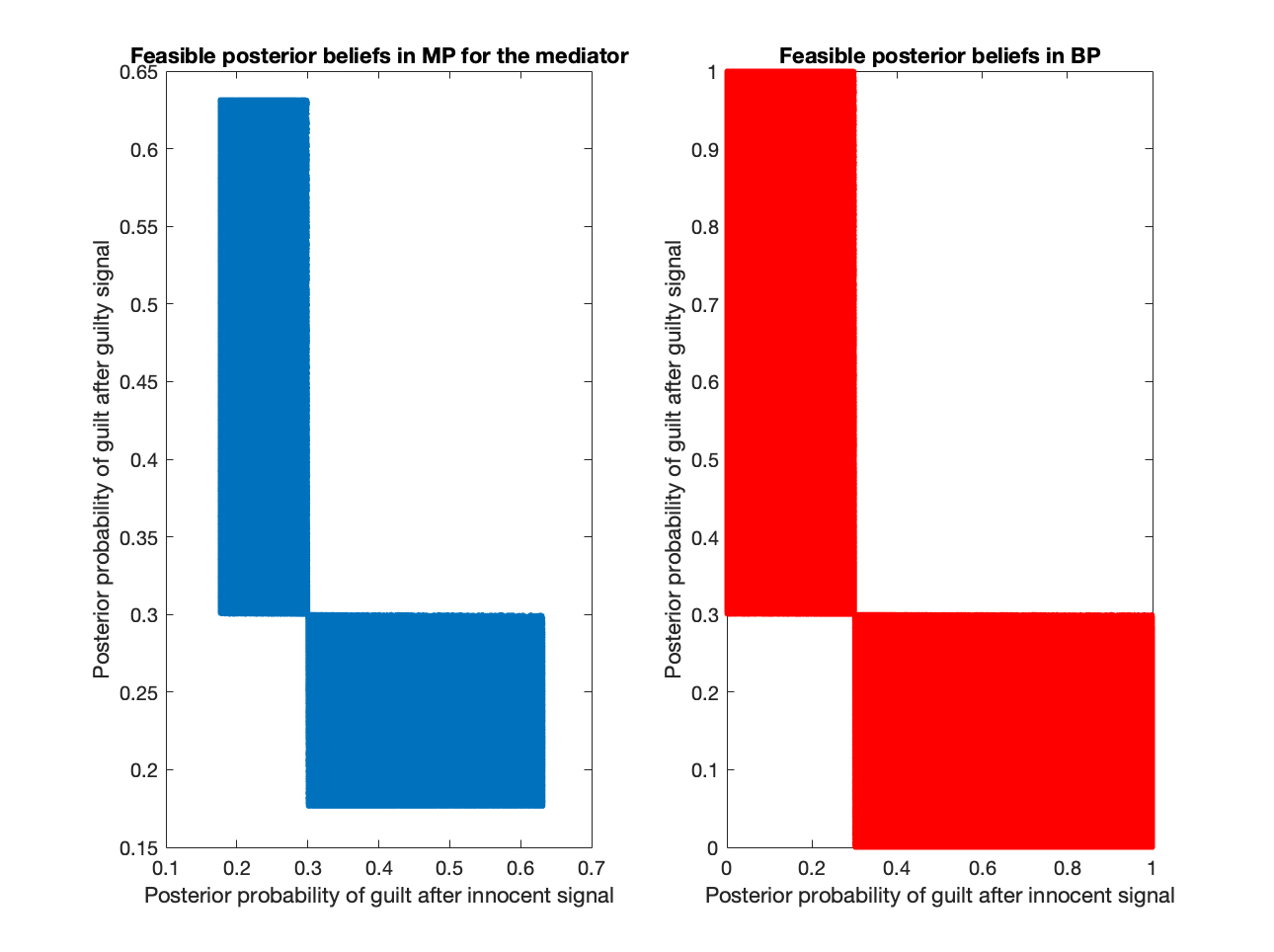}};
\node at (5.2,8) [scale=1.2] {$X=\begin{pmatrix}\frac{6}{7}&\frac{3}{7} \\ \frac{1}{7} &\frac{4}{7} \end{pmatrix}$};
\end{tikzpicture}
\caption{Feasible Set for the Mediator.}
\end{figure}

For a more illustrative example, suppose that $\Omega = \{\omega_0, \omega_1 \}, S=\{s_0, s_1\}, E=\{e_0,e_1\}$ and $A=[0,1]$; we can illustrate the interplay of the choices of the mediator and the sender in figure 4.



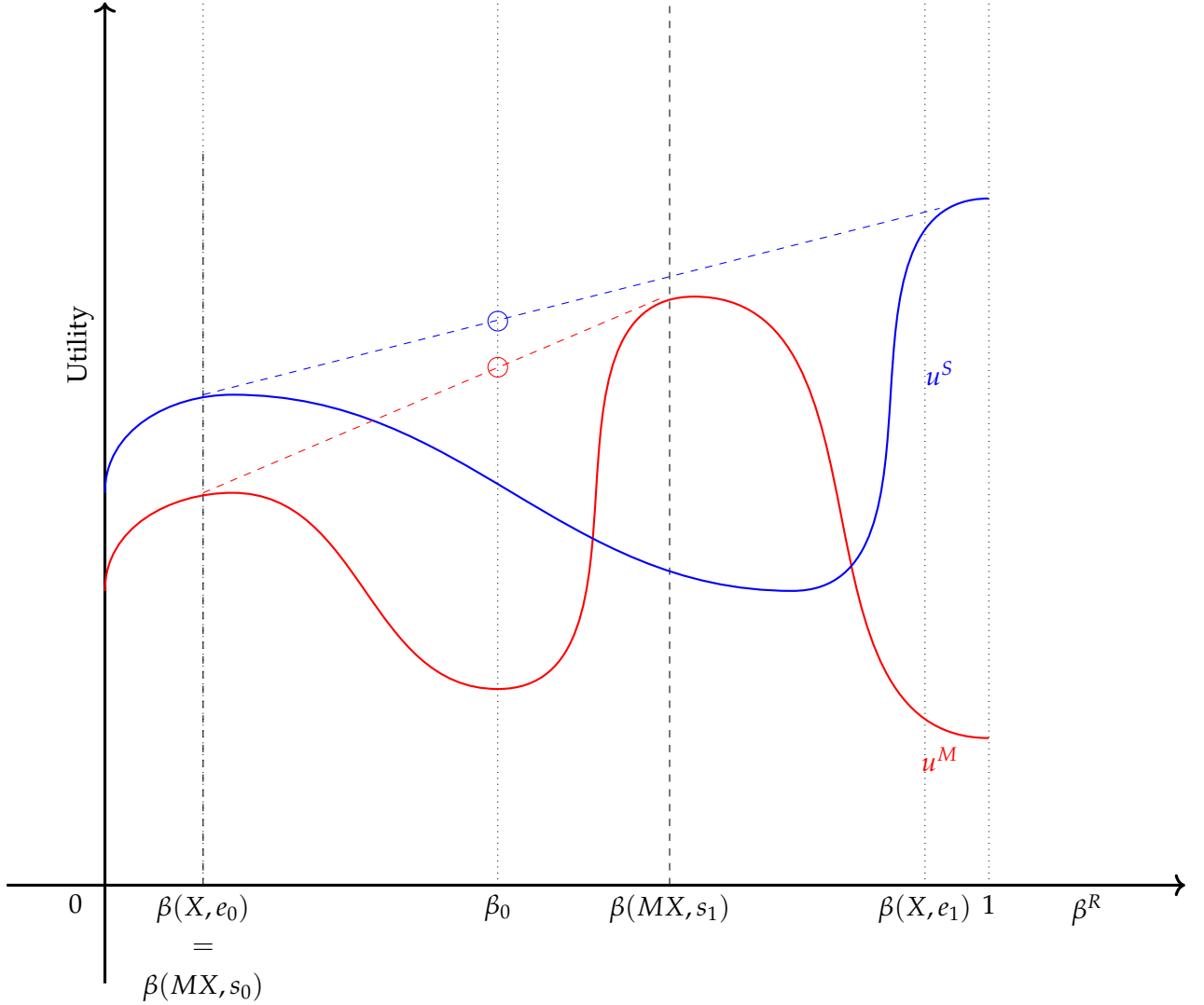
\begin{figure}[h]
\begin{tikzpicture}[scale=1.4]
\draw[very thick,->] (-1,0)--(11,0);
\draw[very thick,->] (0,-1)--(0,9);
\node [below] at (10,0) {$\beta^R$};
\node [below] at (9,0) {1};
\node [below] at (-0.3,0) {0};
\node [below] at (4,0) {$\beta_0$};
\draw[dotted] (4,0)--(4,9);
\node [below] at (1,0) {$\beta(X,e_0)$};
\node [below] at (1,-0.5) {$=$};
\node [below] at (1,-0.8) {$\beta(M X,s_0)$};
\draw [dashed] (1,0)--(1,7.5);
\draw [dotted] (1,0)--(1,9);
\node [below] at (8.35,0) {$\beta(X, e_1)$};
\node [below] at (5.75,0) {$\beta(M X, s_1)$};
\draw [dashed] (5.75,0)--(5.75,9);
\draw [dotted] (8.35,0)--(8.35,9);
\draw [thick,red] (0,3) to [out=90,in=180] (1.3,4)
to [out=0,in=180] (4,2) to [out=0,in=-180] (6,6) to [out=0,in=-180] (9,1.5);
\draw [thick,blue] (0,4) to [out=90,in=180] (1.3,5)
to [out=0,in=180] (7,3) to [out=0,in=-180] (9,7);
\draw [dotted] (9,0)--(9,9);
\node [above,blue] at (8.5,5) {$u^S$};
\node [below,red] at (8.5,1.5) {$u^M$};
\node [left,rotate=90] at (-0.25,6) {Utility};
\draw [dashed,blue] (1,5)--(8.5,6.9);
\draw [blue] (4,5.75) circle [radius=0.1];
\draw [dashed,red] (1,4)--(5.7,6);
\draw [red] (4,5.28) circle [radius=0.1];
\end{tikzpicture}
\caption{An Example.}
\end{figure}

In figure 4, in the absence of a mediator, the sender would concavify her beliefs over the entire belief space and choose the best Bayes-plausible combination, depicted in the figure by $X$ and the two realizations, $e_0$ and $e_1$. However, given that strategy of the sender, the mediator now has an incentive to concavify beliefs over the interval between $\beta(X,e_0)$ and $\beta(X,e_1)$; as drawn she would prefer to keep the left belief where it was and shift the right belief inward; this yield a much higher level of utility. However, note that the sender is now much worse off (and in fact, may be even worse off than she would be had she chosen the babbling experiment in the first place). Now the sender has an incentive to change her action and provide more information by spreading beliefs outward; this kind of interplay is exactly what we focus on. 



 One can also view the signal choice as a (possibly stochastic) \textit{recommendation} from the mediator; this would be particularly convenient if one could identify the signal realization space with the action space. 
 The receiver observes the choices of both the experiment (by the sender) and the signal (by the mediator). This view would be akin to the literature on information design, and thus the sender would be designing an experiment subject to an obedience requirement. This however, is somewhat different from our setting. 
 
For now we focus on the case of a single mediator, as it's the simplest, builds intuition and corresponds most closely with the motivating example.

\begin{subsection}{Building Intuition: A Benchmark With KG Utilities}

One useful illustration of the present model is to compare the outcomes of a particular case of the mediated persuasion model to the leading example of the Bayesian persuasion model presented in KG; doing so also provides a good benchmark for the possible outcomes and builds intuition. To that end, suppose that we take the simple model presented in KG, keep the preferences the same and the add a mediator.  
$\Omega=\{guilty,innocent\},E=S=\{g,i\}$ and $A=\{convict,acquit\}$, let

\begin{equation}
u^S(a)=\begin{cases} 1\mbox{ if } a^R=convict\\
0\mbox{ otherwise}
\end{cases}
\end{equation}
 
and 
 
\begin{equation}
u^R(a,\omega)=
\begin{cases}
1\mbox{ if } \omega=guilty \mbox{  } \& \mbox{  }  a^R=convict\\
1 \mbox{ if }  \omega=innocent \mbox{  } \& \mbox{  }  a^R=acquit\\
0 \mbox{ otherwise }
\end{cases}
\end{equation}
Suppose that the common prior belief of $\omega=guilty$ is $\beta_0=0.3$.
We are left of course, with the question of what the preferences of the mediator are; one of our questions of interest is how do the outcomes vary when we change the mediator's preferences. For this reason we first consider the "extreme" cases - two cases where the mediators preferences coincide with those of the other two players

Case 1: $u^M=u^S$. In this case the interests of the sender and mediator coincide, and clearly, the optimal choice in the Bayesian persuasion model continues to be optimal in the mediated persuasion model. It can be implemented by choosing the same experiment as in the BP model, namely, 
$X= \begin{blockarray}{ccc}
& innocent & guilty  \\
\begin{block}{c(cc)}  
innocent & \frac{4}{7} & 0 \\
guilty & \frac{3}{7} & 1\\ 
\end{block} 
\end{blockarray}$, 
and $\Sigma= \begin{blockarray}{ccc}
& i & g  \\
\begin{block}{c(cc)}  
i & 1& 0 \\
g & 0 & 1\\ 
\end{block} 
\end{blockarray}$. The product $\Sigma X$ would then clearly yield the desired distribution of signals, and the resulting optimal distribution of beliefs. For convenience we reproduce the picture from KG in figure 5:

\begin{figure}
\begin{tikzpicture}[scale=0.75]
\draw[very thick,->] (-1,0)--(11,0);
\draw[very thick,->] (0,-1)--(0,9);
\node [below] at (3,0) {$\beta_0$};
\node [below] at (5,0) {$\frac{1}{2}$};
\draw[thick,dashed] (0,0.1)--(5,0.1);
\draw[thick,dashed] (5,7)--(10,7);
\draw[thick,dotted] (0,0)--(5,7)--(10,7);
\node [left] at (0,7) {1};
\draw (-0.1,7)--(0.1,7);
\node [below] at (10,0) {1};
\draw (10,0.1)--(10,-0.1);
\end{tikzpicture}
\caption{The KG Setting.}
\end{figure}
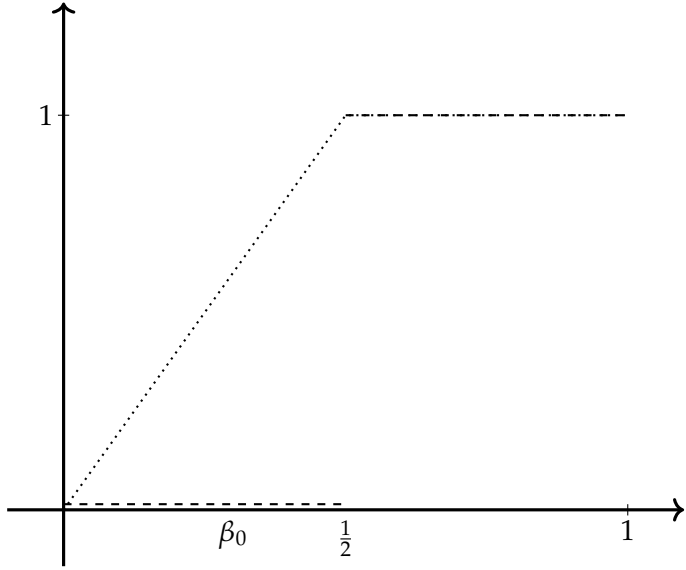

The $X$ and $\Sigma$ above do not constitute, however, a unique equilibrium. In fact, any pair $(\Sigma,X)$ with the property that their product results in a Bayes-plausible combination of the beliefs $\beta=0$ and $\beta=0.5$ is an equilibrium. This simple example shows that the mere presence of a mediator can increase the number of equilibria, but keep the outcome the same. 

Case 2: $u^M=u^R$. We now turn to the question of what happens if the mediator's preferences are fully aligned with those of the receiver. While intuition suggests that this arrangement is must be better for the receiver, we show by example that in fact, this does not have to be strictly so.\footnote{In fact, later we show that more generally, the receiver benefits from persuasion when the mediator's preferences are \textit{not} fully aligned with those of the receiver.} Writing the mediator's utility as a function of the receiver's belief we obtain

$u^M(\beta)=
\begin{cases}
1-\beta \mbox { if } \beta <\frac{1}{2}\\
\beta \mbox{ if } \beta \geq \frac{1}{2}
\end{cases}
$

which we plot on figure 6 in red.

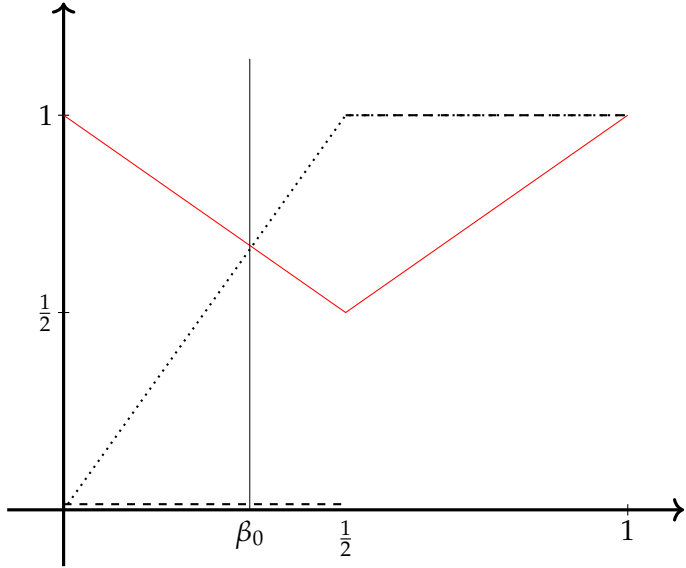
\begin{figure}
\begin{tikzpicture}[scale=0.75]
\draw[very thick,->] (-1,0)--(11,0);
\draw[very thick,->] (0,-1)--(0,9);
\node [below] at (3.3,0) {$\beta_0$};
\node [below] at (5,0) {$\frac{1}{2}$};
\draw[thick,dashed] (0,0.1)--(5,0.1);
\draw[thick,dashed] (5,7)--(10,7);
\draw[thick,dotted] (0,0)--(5,7)--(10,7);
\node [left] at (0,7) {1};
\draw (-0.1,7)--(0.1,7);
\node [below] at (10,0) {1};
\draw (10,0.1)--(10,-0.1);
\draw[red] (0,7)--(5,3.5)--(10,7); 
\draw (-0.1,3.5)--(0.1,3.5);
\node [left] at (0,3.5) {$\frac{1}{2}$};
\draw[ultra thin] (3.3,0)--(3.3,8);
\end{tikzpicture}
\caption{Adding a Mediator with $u^S=u^M$ to KG.}
\end{figure}
\end{subsection}

The concavification of $u^M$ over the entire belief space (which we do not depict) is simply a straight line at 1. If the sender induces the same two beliefs ($\beta=0$ and $\beta=0.5$) as in the base case, since any garbling of these two beliefs would induce beliefs that are interior to the set $[0,0.5]$ and since the mediator's utility is linear in the subset of belief space that is feasible (and therefore the constrained concavification coincides with utility everywhere), the mediator is indifferent between any Bayes-plausible garbling of the two beliefs. As for the sender, she gets zero utility from any beliefs $\beta \in [0,\frac{1}{2})$. Since the mediator is indifferent over the space of constrained beliefs, in particular, the original equilibrium outcome can be sustained in the same way as above - the sender plays $X$ and the mediator truthfully reproduces the experiment realization. 

Observe however, that if the mediator were to play any nontrivial garbling, that would no longer be an equilibrium, since then the sender would get utility zero (as opposed to getting 0.6 in equilibrium), and would have an incentive to "undo" the garbling, bringing the beliefs back outward. Additionally, it is also not an equilibrium for the sender to play something that is strictly more informative than $X$, since then one of the beliefs would be above $\frac{1}{2}$, in which case the mediator's utility would be convex over the set of possible posterior beliefs, and the mediator would have a strict incentive to play a fully revealing $\Sigma$, in which case the sender would prefer to deviate back to the $X$ described above. 

Suppose that the sender chooses a particular experiment $X$ and the mediator chooses a particular experiment $\Sigma$. Observe that then the receiver is computing the posterior belief from a combined distribution that is simply the product of the two choices: $\Sigma X \triangleq B$. Since $\Sigma$ is a column-stochastic matrix, as noted above, this is precisely the definition of $B$ being Blackwell-inferior (\hyperlink{Blackwell (1951)}{Blackwell (1951}), \hyperlink{Blackwell (1953)}{Blackwell (1953)}) to $X$ with $\Sigma$ being the garbling matrix. Thus, \textit{whatever} the mediator chooses, the resulting distribution of signal realizations will be dominated by the sender's experiment in the sense of Blackwell. Blackwell's characterizations immediately apply and we have the following series of results which we state without proof since they are direct consequences of Blackwell's theorem. 

\begin{observation}
The distribution of receiver beliefs under $X$ is a mean-preserving spread of the distribution of receiver beliefs under $B$.
\end{observation}

It is immediate that if the sender and the mediator have the same preferences, full revelation may not be an equilibrium (in that case the set of nontrivial equilibrium outcomes coincides with that in KG). In \hyperlink{Gentzkow and Kamenica (2017a)}{Gentzkow and Kamenica (2017a)} and \hyperlink{Gentzkow and Kamenica (2017b)}{Gentzkow and Kamenica (2017b)} full revelation is typically an equilibrium (with at least two senders); the reason is that they identify a condition on the informational environment ("Blackwell-connectedness") which guarantees that each player can unilaterally deviate to a Blackwell-more informative outcome, regardless of the actions of the other player. Preference divergence then forces full revelation. Finally, adding senders does not make the uninformative equilibrium disappear. 

\begin{section}{Binary Model}

For tractability we work with a binary model where there are two states of the world and two experiment and signal realizations. This is with (perhaps significant) loss of generality, but will serve well to illustrate the basic idea of how to compute a best response for the sender given the choice of the mediator. 

\begin{subsection}{Computing the Set of Feasible Posteriors}

Setting aside the issues of strategic behavior for now, we first ask a simpler question: given a \textit{fixed}\footnote{I.e. not strategically chosen by a player as a function of her preferences.} signal (or equivalently, a fixed garbling), or a fixed experiment, what are all the posterior distributions that can be induced? At this point we can make an important connection with the cheap talk and communication literature. \hyperlink{Blume, Board and Kawamura (2007)}{Blume, Board and Kawamura (2007)} discuss a model of cheap talk where the signal sent by the sender is subject to random error - with a small probability the message observed by the receiver is not the message sent by the sender, but rather, a message sent from some other distribution that does not depend on the sender's type or the message chosen. We make this connection to note that choosing an information structure that will be subjected to a fixed, non-strategically-chosen garbling is exactly equivalent to choosing a random signal that will be subject to noise. Thus, our model subsumes a model on Bayesian persuasion with noisy communication, similar to those studied by \hyperlink{Le Treust and Tomala (2018)}{Le Treust and Tomala (2018)} and \hyperlink{Tsakas and Tsakas (2018)}{Tsakas and Tsakas (2018)}.

In the (different but related) setting of cheap talk, as noted by \hyperlink{Ambrus, Azevedo and Kamada (2013)}{Ambrus, Azevedo and Kamada (2013)} as well as \hyperlink{Blume, Board and Kawamura (2007)}{Blume, Board and Kawamura (2007)} stochastic reports make incentive compatibility constraints \textit{easier} to satisfy. This will not quite be the case here, but this will nevertheless be an illuminating exercise. 

As mentioned above, for tractability\footnote{And with loss of generality, which we discuss later.} we will work in the simplest possible environment of binary signal and state spaces for both the sender and the mediator. In addition to being the simplest nontrivial example of the problem we are trying to solve, working with two-by-two square matrices has a very important additional advantage. The rank of such a stochastic\footnote{Which of course, rules out the zero matrix, which has rank zero.} matrix can be only two things - one or two. If the rank of a two-by-two stochastic matrix is one, that means that not only the columns (and rows) are linearly dependent, but they must, in fact be identical. In that case the garbling is fully uninformative - it can be readily checked that this results in the same posteriors as the canonical complete garbling; namely, the posterior (after either signal realization) is equal to the prior. The other possible case is that the matrix has rank two - but that automatically means that such a matrix is invertible. We shall use this fact of existence of an inverse extensively\footnote{We also comment on the interpretation of the rank of a garbling matrix later in the discussion, and in related contemporaneous work}.  

More specifically, let $\epsilon$ be a small positive number, set the space of experiment realizations to be $E=\{e_L,e_H \}$ and suppose that the sender and receiver play a game exactly identical to KG (that is, there is no mediator), except that with probability $\epsilon$ the signal observed by the receiver (denoted by $e^o$) is not the signal sent (which we denote by $e^s$), but a signal chosen from the following distribution 
$e^o=
\begin{cases} 
e_H & \text{with probability }  p\\
e_L & \text{with probability }  1-p
\end{cases}$ 

The key thing is that this distribution is independent of both the type and the signal realized. Thus, we can compute the probabilities of observed signals as functions of the parameters and realized signals as usual:

\begin{equation}
\mathbb{P}(e^o=e_H|e^s=e_H)=1-\epsilon +\epsilon p
\end{equation}
\begin{equation}
\mathbb{P}(e^o=e_L|e^s=e_H)=\epsilon -\epsilon p
\end{equation}
\begin{equation}
\mathbb{P}(e^o=e_L|e^s=e_L)=1-\epsilon p 
\end{equation}
\begin{equation}
\mathbb{P}(e^o=e_H|e^s=e_L)=\epsilon p 
\end{equation}

Then this is equivalent to having a garbling 
\begin{equation}
\Sigma=\begin{pmatrix} \sigma_1 & \sigma_2\\1-\sigma_1 & 1-\sigma_2\end{pmatrix}=
\begin{pmatrix} \epsilon p - \epsilon + 1 &   \epsilon p\\\epsilon - \epsilon p & 1 - \epsilon p  \end{pmatrix}
\end{equation}
 with realization space $S=\{e^o_L, e^o_H\} $.

If we denote by $X=\begin{pmatrix}x&y\\1-x&1-y \end{pmatrix}$ the experiment chosen by the sender so that 
\begin{equation}
B=\Sigma X=
\begin{pmatrix} x(\epsilon p - \epsilon + 1) - \epsilon p(x - 1)& y(\epsilon p - \epsilon + 1) - \epsilon p(y - 1)\\
(\epsilon p - 1)(x - 1) + x(\epsilon - \epsilon p) & (\epsilon p - 1)(y - 1) + y(\epsilon - \epsilon p)   
\end{pmatrix}
\end{equation}
 is the resulting distribution of signal observations given states. 
Letting $\Omega=\{ \omega_H,\omega_L\}$ be the set of states and setting prior belief of $\omega_L=\pi$ the posterior beliefs are
\begin{equation}
\beta(s_H)=\mathbb{P}(\omega_L |s_H)=\frac{\pi \left[y(\epsilon p - \epsilon + 1) - \epsilon p(y - 1) \right]}{\pi \left[y(\epsilon p - \epsilon + 1) - \epsilon p(y - 1) \right]+(1-\pi)\left[x(\epsilon p - \epsilon + 1) - \epsilon p(x - 1) \right]}
\end{equation}
 and 
\begin{equation} 
\beta(s_L)=\mathbb{P}(\omega_L |s_L)=\frac{\pi \left[ (\epsilon p - 1)(y - 1) + y(\epsilon - \epsilon p) \right]}{\pi \left[ (\epsilon p - 1)(y - 1) + y(\epsilon - \epsilon p) \right]+ (1-\pi)\left[ (\epsilon p - 1)(x - 1) + x(\epsilon - \epsilon p)\right]}
\end{equation}

Define the set of feasible beliefs to be a pair 
\begin{equation}
F(M,\pi)\triangleq \{(\beta(s_H),\beta(s_L)\in [0,1]^2 ) | \beta(s_H),\beta(s_L) \in \supp (\tau(MX)), \exists X \in \mathbf{X} \} 
\end{equation}
 One observation we can immediately make is that the set of feasible beliefs with a garbling is a strict subset of the set of feasible beliefs without one, simply due to the fact that there are extra restrictions in computing $F(M,\pi)$. To illustrate, let $\epsilon=\frac{1}{100}$ and $p=\frac{1}{4}$ so that there is a 1\% chance that the signal will be a noise signal, and if that happens, there is a 75\% probability that the signal will be correct. The set of Bayes-plausible beliefs is depicted in red in the figure 7, while the set of feasible beliefs given this particular $\Sigma$ is in blue. 

\begin{figure}{}
\includegraphics[scale=0.35]{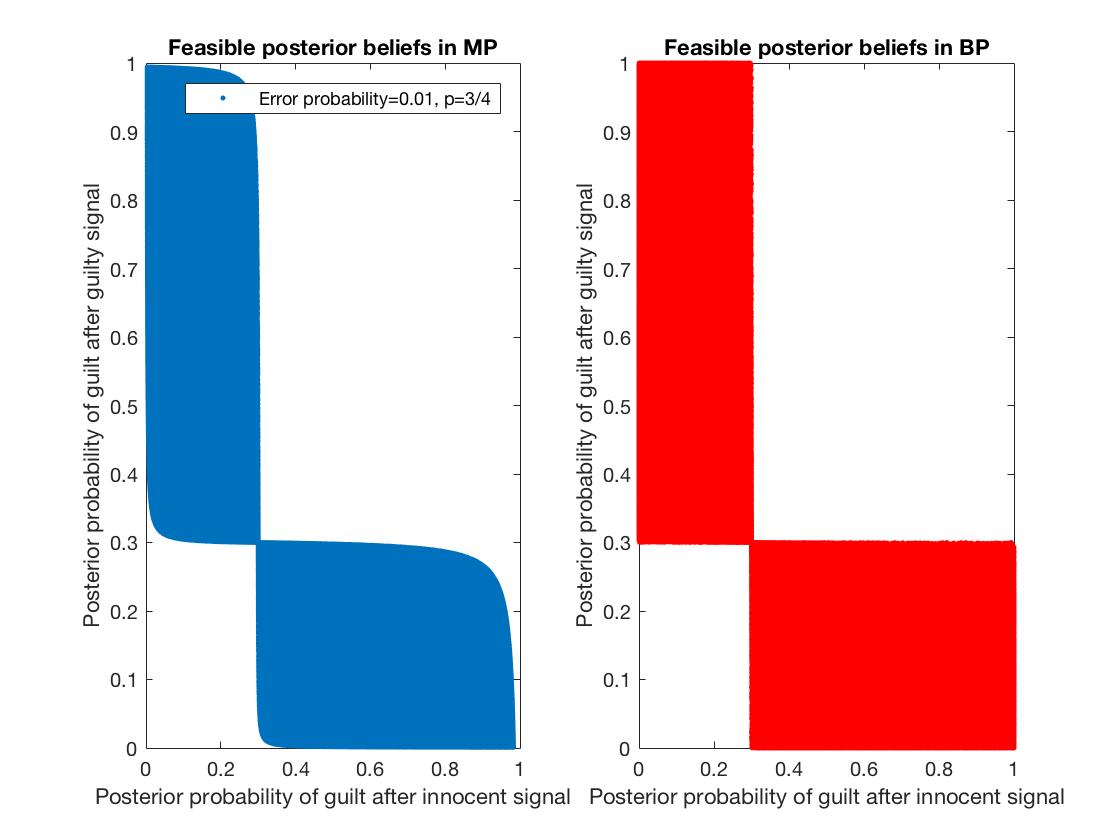}
\caption{Comparing the Feasible Sets of Posteriors.}
\end{figure}

Clearly the "butterfly" set of feasible beliefs (left) is a strict subset of the Bayes-plausible set on the right, verifying the observation made above. Thus, \textit{for a fixed garbling, not all Bayes-plausible posterior beliefs can be induced.}

Perhaps another illustration can make this point more starkly - suppose we were to increase the probability of error tenfold, so that there is a much greater chance that the signal is a noise signal. The resulting sets are depicted in figure 8.
\begin{figure}
\includegraphics[scale=0.35]{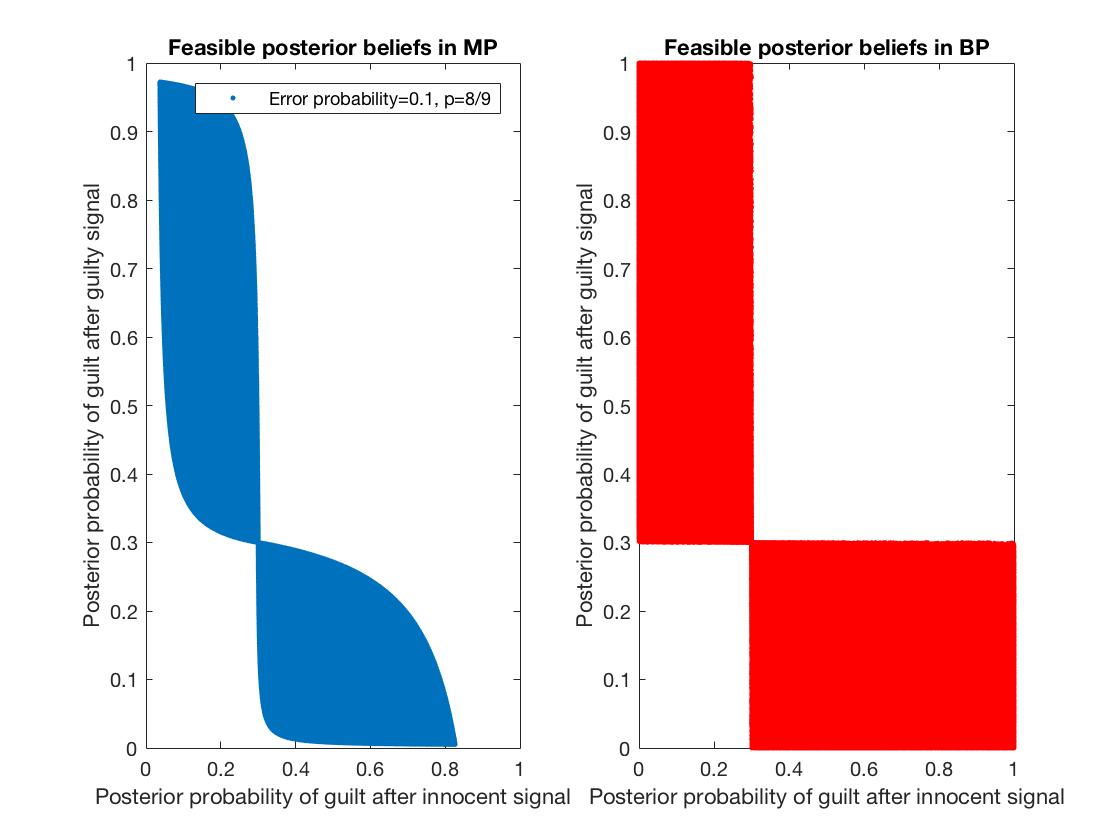}
\caption{Increasing Noise Shrinks the Set of Feasible Posteriors.}
\end{figure}

Thus, increasing the probability of error (or noise signal) shrinks the set of feasible beliefs. This is consistent with intuition - if the signal is pure noise, then there should not be any update of beliefs (and thus the set would shrink to a single point at the prior), and with a larger probability of noise one would update "less". We make precise the idea that with a less informative garbling "fewer" posteriors are available below. 

This discussion leads to the following question: What is the set of feasible posterior beliefs given a garbling (without computing whether or each belief is feasible one by one as was done in computing the figures above, which were generated by simulating random matrices with the appropriate stochasticity constraints)? One way of answering this question is to trace out the confines of the feasible set. As luck would have it, there is an observation we can make that simplifies this a great deal. If we fix one posterior belief (say, $\beta_1$ the posterior after the innocent signal) and then ask what would the elements $X$ need to be to either maximize or minimize the other posterior belief, it turns out that either $x$ or $y$ (or both) will always be 1 or 0.
We fix $\Sigma = \begin{pmatrix}\sigma_1&\sigma_2\\1-\sigma_1&1-\sigma_2 \end{pmatrix}$, let $\pi$ be the prior belief and consider $X=\begin{pmatrix}x&y\\1-x&1-y \end{pmatrix}$.
Computing outer limits of $F(\Sigma,\pi)$ is equivalent to the following program:
\begin{equation}
\max_{x,y} \beta_2=\frac{\pi[\sigma_1 y + \sigma_2(y-1) ]}{\pi[\sigma_1 y + \sigma_2(y-1) ] +(1-\pi)[\sigma_1 x - \sigma_2 (x-1) ]}
\end{equation}
\begin{equation}
s.t.\hspace{0.5cm} \beta_1=const.
\end{equation}
\begin{equation}
0\leq x \leq 1; 0\leq y \leq 1
\end{equation}
The solution (which we do not exhibit, as it is straightforward but somewhat tedious) shows that either $x$, or $y$ or both will be 0 or 1 (and of course, we could also have fixed $\beta_2$ and let that be the parameter; the answer would be the same). The result is intuitive (maximizing a posterior belief requires maximizing the probability of one of the signals in the first place), but this verifies the intuition formally.

Again, fortunately for us, this observation can be operationalized in the following way: we first fix one of four extreme points of the $X$ matrix, and then trace out the corresponding possible beliefs by systematically varying the other probabilities in the experiment, which yields a  curve (or a path, in topological terms) parametrized by a single number - the probability of one of the signals. 

We illustrate this approach using $M = \begin{pmatrix} \frac{1}{3} & \frac{1}{7}\\ \frac{2}{3} & \frac{6}{7} \end{pmatrix}$. The question is, what is $F(\Sigma,\pi)$ for this garbling? We use the algorithm just prescribed: first fix a perfectly revealing part of the experiment, and then vary the corresponding distribution. 

Letting $X^1=\begin{pmatrix} 1 & p\\ 0 & 1-p \end{pmatrix}$ and varying $p$ from 0 to 1 yields the following (blue) curve in figure 9.

\begin{figure}
\includegraphics[scale=0.35]{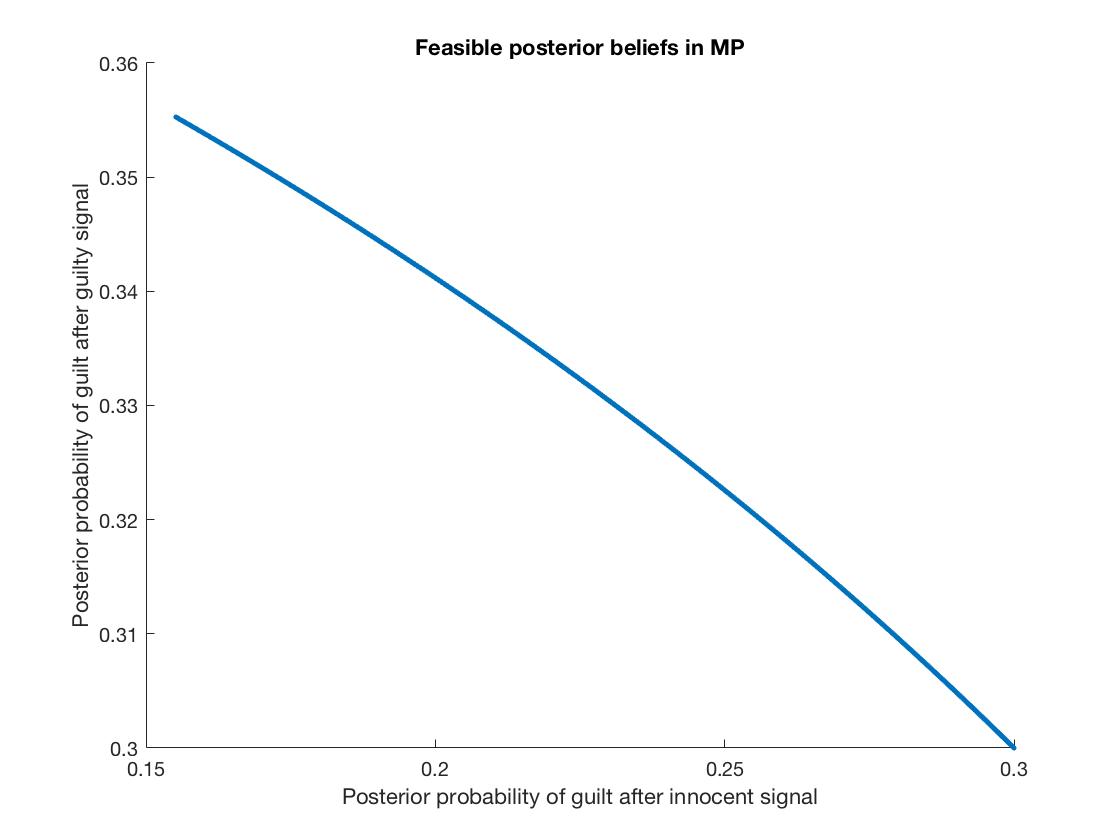}
\caption{Tracing the Outer Limit of $F(\Sigma,\pi)$: First Boundary.}
\end{figure}

Now we fix the next extreme point: $X^2=\begin{pmatrix} 0 & p\\ 1 & 1-p \end{pmatrix}$ and again vary $p$, which yields the following (reddish-brown) boundary in figure 10.

\begin{figure}
\includegraphics[scale=0.35]{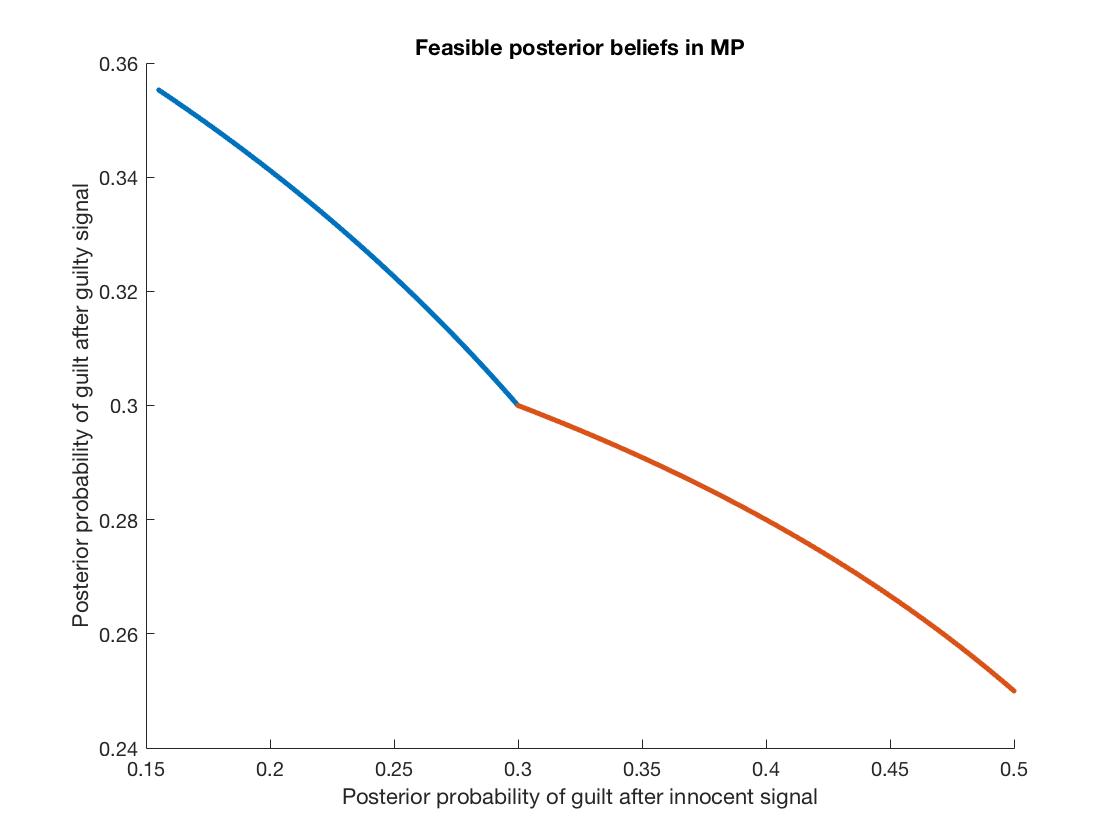}
\caption{Tracing the Outer Limit of $F(\Sigma,\pi)$: Second Boundary.}
\end{figure}

Next we fix the third extreme point: $X^3=\begin{pmatrix} p & 1\\ 1-p & 0 \end{pmatrix}$ and trace the corresponding (yellow) curve, illustrated in figure 11.

\begin{figure}
\includegraphics[scale=0.35]{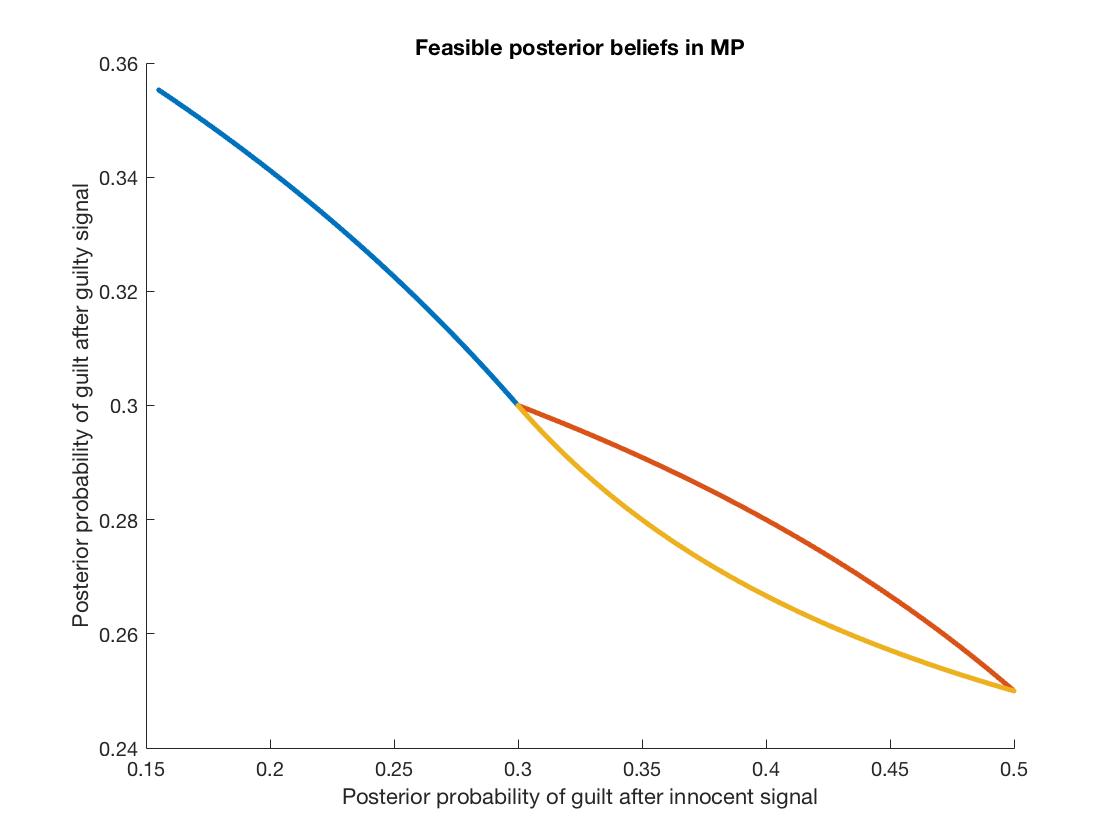}
\caption{Tracing the Outer Limit of $F(Sigma,\pi)$: Third Boundary.}
\end{figure}

And finally we trace out the last (purple) curve by using $X^4=\begin{pmatrix} p & 0\\ 1-p & 1 \end{pmatrix}$ in figure 12.
 
\begin{figure}
\includegraphics[scale=0.35]{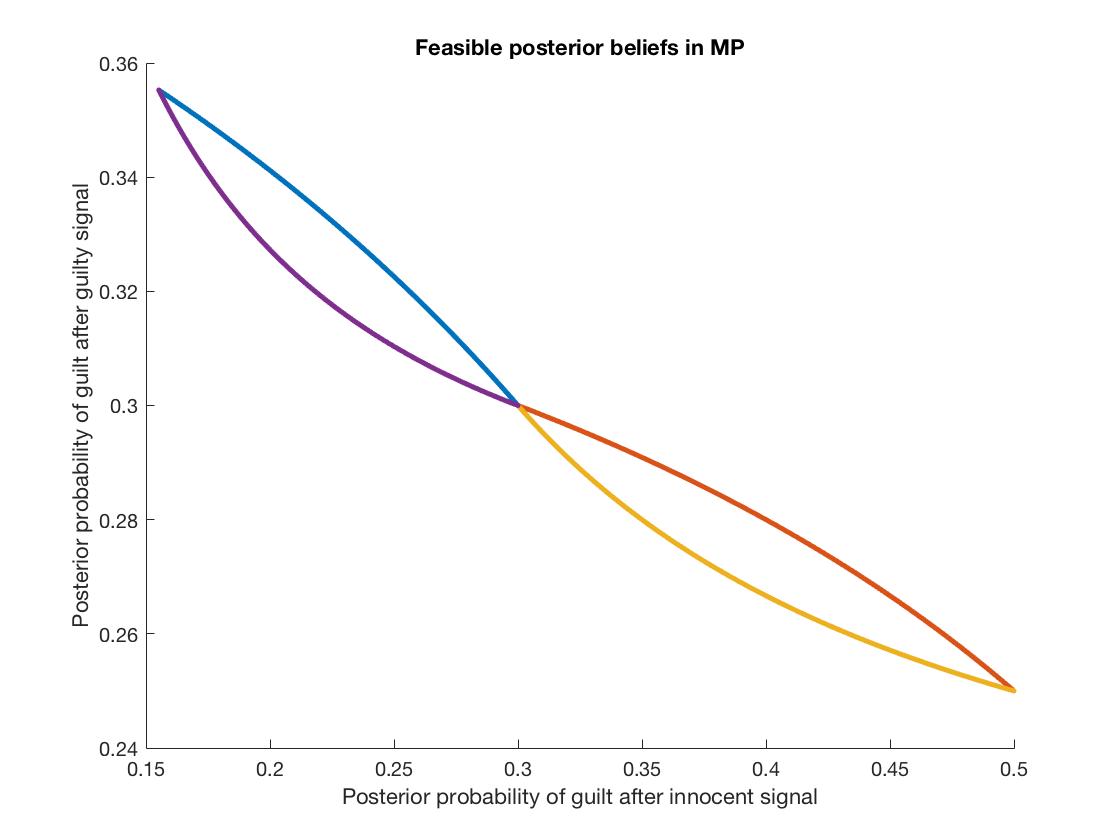}
\caption{Tracing the Outer Limit of $F(\Sigma,\pi)$: Fourth Boundary.}
\end{figure}

\begin{figure}
\includegraphics[scale=0.35]{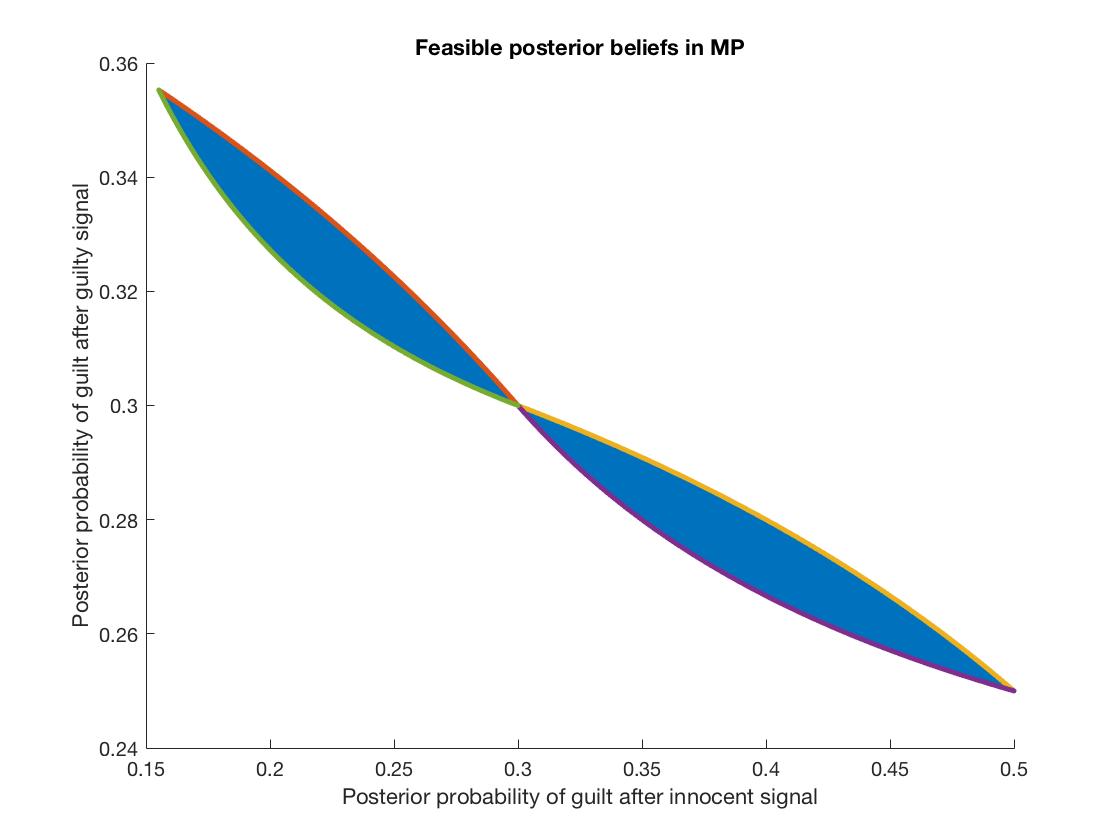}
\caption{$F(\Sigma,\pi)$: an Illustration.}
\end{figure}

This procedure is a simple way of computing the set of  $F(\Sigma,\pi)$; \textit{this procedure is a complete characterization of the set of feasible beliefs for $2\times2$ signals and experiments}. Now, for a belief in this set we can ask: does there exist an experiment that yields this belief, and if so, how do we compute it?

One of the implications of Proposition 1 in KG is that for every Bayes-plausible posterior distribution there exists an experiment that induces that distribution; they also give an explicit formula for computing such an experiment. In mediated persuasion this fails - an experiment inducing a particular Bayes-plausible distribution may not exist, if it is garbled. However, for beliefs that are feasible given $\Sigma$ we have a simple formula for computing the experiment that induces those beliefs. 

\begin{defi}
Fix $\Sigma$. A distribution of posterior beliefs $\tau$ is said to be $\Sigma$-plausible if there exists a stochastic matrix $X$ such that $p(\Sigma X)=\tau$.
\end{defi}

\begin{theorem}
Fix $\Sigma$. Suppose that $\tau$ is a Bayes-plausible and $\Sigma$-feasible distribution of posterior beliefs. There exists an experiment $X$ such that $p(\Sigma X)=p(B)=\tau$.
\end{theorem}

We construct the entries in B by setting $b(s|\omega)=\frac{\beta(\omega|s )\tau(\beta)}{\pi(\omega)}$ as in KG; simple algebra shows that this yields a Bayes-plausible distribution that results in the necessary beliefs. The experiment yielding $B$ is then simply $X=\Sigma^{-1}B$\footnote{Note that we are using the existence of $\Sigma^{-1}$}. The fact that $X$ is, in fact, an experiment is guaranteed by the fact that the beliefs were feasible in the first place. This is, in a sense, a tautological statement, but it does provide an analogue to Proposition 1 in KG by exhibiting an explicit formula for constructing $B$ and then $X$ and showing that both do, in fact, exist. 

The above example and proposition suggest a general way of solving the problem with two states, two signal realizations and two experiment realizations with a fixed garbling $\Sigma$. First we compute the four outer limits of $F(\Sigma,\pi)$ as above. Then we ask how the sender's utility varies over the feasible set, and having found a maximum point, we construct the optimal experiment using theorem 3.1. And then, given the feasible set of a garbling, one can compute the sender's utility from choosing each posterior in that set (simply plot the sender's utility as a function of the posterior beliefs), find the maximal beliefs and construct the experiment yielding those beliefs. \textit{This procedure shows how to find a best response for the sender.}

We can write this problem and its solution more formally, which we do now. Let $\kappa$ be the constant and denote the maximization program by P. Suppose that the program has a solution\footnote{This amounts to assuming that the $\kappa$ can actually be a posterior belief, which is not always the case - take for example the belief $\beta_1=0.9$ in figure 6. Such a belief is clearly infeasible for that $\Sigma$, and thus the program would not have a solution.} and denote by $x^*(\sigma_1,\sigma_2,\pi,\kappa)$ the solution. Suppose for now that $\kappa \leq \pi$. This produces a (second posterior belief) function $\beta_2^{max}(y;x^*(\sigma_1,\sigma_2,\pi,\kappa),\sigma_1,\sigma_2,\pi):[0,1]\rightarrow[0,1]$ we write it to emphasize that all arguments of the $\beta^{max}_2$ function after the semicolon are parameters, and only the $y$ argument is varying from $0$ to $1$. Analogously we can compute $\beta_2^{min}(y;x^*(\sigma_1,\sigma_2,\pi,\kappa),\sigma_1,\sigma_2,\pi):[0,1]\rightarrow[0,1]$. Let $Gr(\beta^{max}_2)$ and $Gr(\beta^{min}_2)$ be the graphs of the two functions, and let $Co(A)$ be the convex hull of an arbitrary nonempty set $A$. We then define $F^1(\Sigma,\pi) \triangleq Co(Gr(\beta^{max}_2) \cup Gr(\beta^{min}_2))$; the reason that we can do that is that we have the set of posterior beliefs is convex (because the set of information structures is convex, and Bayes rule is monotonic). Similarly, for $\kappa \geq \pi$ we can compute analogous objects, and define  $F^2(\Sigma,\pi)$. Finally, we let $F(\Sigma,\pi)\triangleq F^1(\Sigma,\pi)\cup F^2(\Sigma,\pi)$.


\begin{figure}[h]
\begin{tikzpicture}

\node[anchor=south west,inner sep=0] at (0,0) {\includegraphics[scale=0.35]{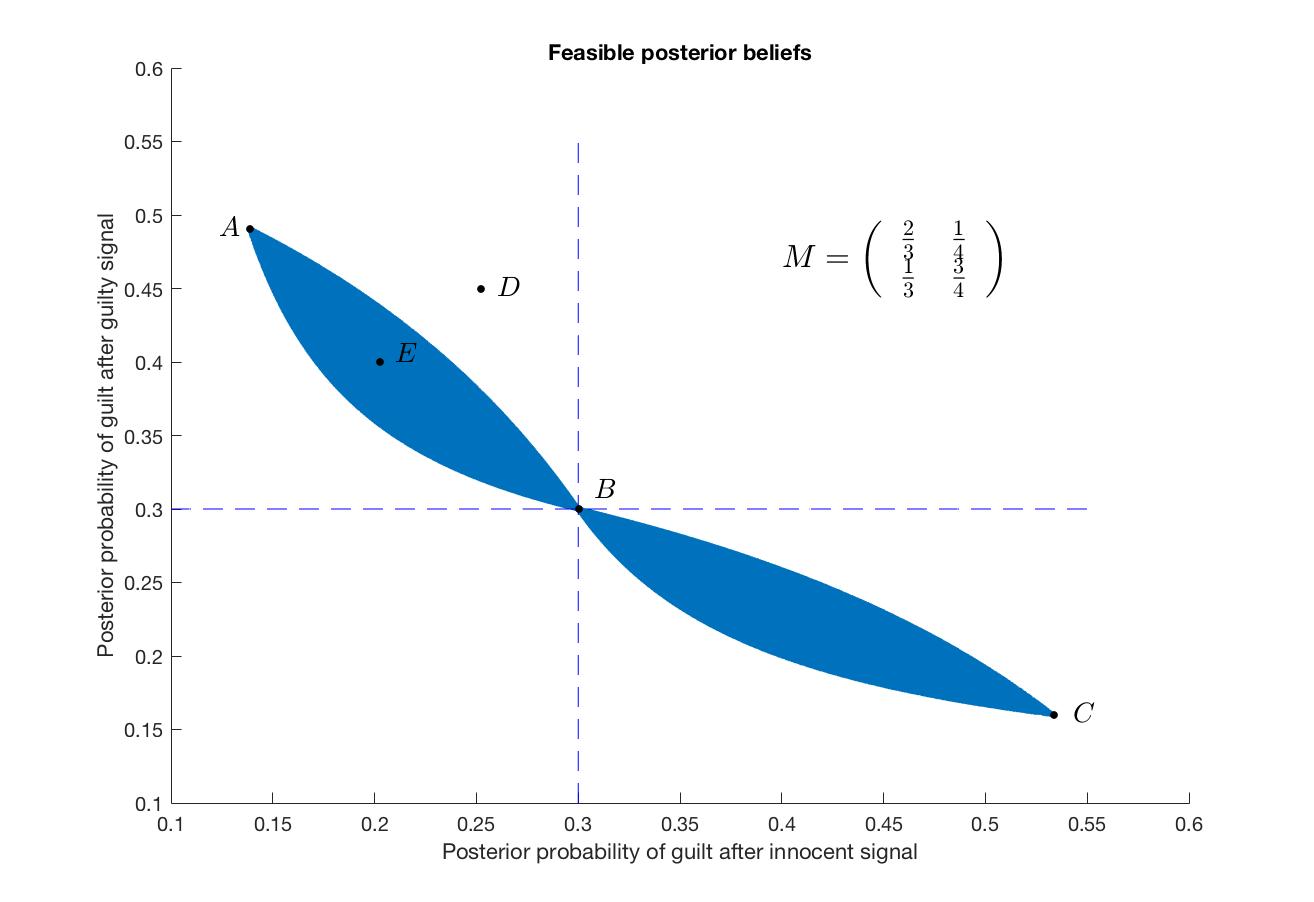}};

\filldraw [fill=white,draw opacity=0] (8,6)--(13,6)--(13,10)--(8,10); 
\node at (10,8) [scale=1.7] {$\Sigma=\begin{pmatrix}\frac{2}{3}&\frac{1}{4} \\ \frac{1}{3} &\frac{3}{4} \end{pmatrix}$};
\end{tikzpicture}
\caption{Major Features of the Feasible Set}
\end{figure}

There are a number of important and interesting observations about the $\Sigma$-feasible set that we can make at this point. Consider the $F$ set illustrated in figure 11, using the garbling matrix $\begin{pmatrix} \frac{2}{3} & \frac{1}{4} \\ \frac{1}{3} & \frac{3}{4} \end{pmatrix}$. In this set each point corresponds to an experiment for the sender. The first thing to notice is that the so-called "butterfly" has two "wings". The "left" wing - the one including point $A$, i.e. the wing up and to the left from the "origin" (i.e. the point where the posteriors are equal to the prior), is the set that would result if the sender were using "natural" signals - i.e. a guilty signal is more likely in the guilty state and an innocent signal is more likely in the innocent state. The right wing is the set that would result if the sender were instead using "perverse" signals - a \textit{guilty} signal that is more likely in the \textit{innocent} state, and vice versa.\footnote{Note that if the sender were to choose a signal, say, guilty, that is more likely in both states, that would quickly bring beliefs back to the prior, and whether it would be in the right or the left wing would be dictated by the relative probabilities.} This is also equivalent to flipping the labels on the signals.

Consider point $B$, the point where both posteriors are equal to the prior (with the obvious motivation, we call that the "origin"). Observe that moving weakly northwest meaning decreasing the first posterior while increasing the second - in other words, a mean-preserving spread.\footnote{The fact that the spread is mean preserving comes from Bayes rule.} Thus, points that are northwest of $B$ are posteriors that are Blackwell-more informative than $B$. Equivalently, they correspond to signals that Blackwell dominate the uninformative signals. Iterating this, point $A$ is Blackwell-most informative among all the points in the left wing. It can also be verified that point $A$ is \textit{precisely} the two posteriors that correspond to the sender using the fully informative (and "natural") signal. The exact opposite logic applies to the right wing, so that $C$ is the extreme posterior corresponding to the Blackwell-most informative "perverse" signal. Importantly, this logic works only within each wing, (or quadrant by quadrant, which are delineated by the dashed lines), and not on the figure as a whole. 

The other observation that we can make is that while $F$ seems symmetric around the "origin", in general, it is not. The lack of symmetry comes from the constraints (and biases) imparted by the garbling; $F(M,\pi)$ is symmetric if and only if $M$ is symmetric.

\begin{defi}
$F$ is said to be \textit{symmetric} if for each $\{\beta_1,\beta_2\}$ if the ordered pair $\{\beta_1,\beta_2\} \in F$ then the ordered pair $\{\beta_2,\beta_1\}$ is also in $F$.
\end{defi}

The next observation is that each wing of the butterfly is convex, but the butterfly itself is not. This comes from the fact that for normal (and respectively, for perverse) signals, if two posteriors can be induced, than so can any convex combination (since the set of the relevant stochastic matrices is convex). On the other hand, for the entire set to be convex, taking a point from the left wing, a point from the right and requiring that a mixture would also be in the set would require each signal to be weakly more likely in either state - which is impossible, except for the degenerate case. This is why we can take the convex hull of the extreme beliefs and outer limits for each wing, but not the convex hull of the entire butterfly. 

The final observation that we can make is the following: the sender is certainly capable of choosing the identity experiment, and inducing $\Sigma I=B$ (in figure 14 this would correspond to point $A$); this is the best (in the sense of being Blackwell-maximal) that the sender can induce. Since the sender can also choose any less informative experiment, it would seem that the sender may be capable of inducing \textit{any} Blackwell-inferior distribution to $A$. Figure 11 shows that this intuition is false. A point like $D$ is certainly Blackwell-inferior to $A$, being a mean-preserving contraction, yet it is outside the feasible set. The question then arises, why can we not simply "construct" the required experiment $X$ as follows: suppose $\Sigma I \succeq B'$ and $p(B')=D$. If there exists an $X$ with $\Sigma X=B'$, we would be done. What about simply putting $X=\Sigma^{-1}B'$? The answer is that \textit{if} $p(\Sigma \Sigma^{-1}B')$ is in $F$, this would work. It turns out that if that it not true, then $\Sigma^{-1}B'$ will not yield a stochastic matrix $X$ and therefore would not be a valid experiment (this can be seen by example). In other words, the sender is not capable of inducing any posterior belief that is Blackwell-inferior to $\Sigma I$.

There are a number of interesting results that we can illustrate using this technique of characterizing the feasible sets. To give but one example, we give a simple proof of a result first described in \hyperlink{Bohnenblust, Shapley and Sherman (1949)}{Bohnenblust, Shapley and Sherman (1949)}, and alluded to in Blackwell's original work (\hyperlink{Blackwell (1951)}{Blackwell (1951)}, \hyperlink{Blackwell (1953)}{Blackwell (1953)}):

\begin{theorem}
Suppose $\Sigma_1$ and $\Sigma_2$ are two garblings with $\Sigma_1 \succeq_{B} \Sigma_2$. Then $F(\Sigma_2,\pi) \subseteq F(\Sigma_1,\pi)$. 
\end{theorem}

\begin{proof}
Fix any $\pi$. We must show that for any $\tau$ if $supp(\tau) \in F(\Sigma_2,\pi)$, then $supp(\tau) \in F(\Sigma_1,\pi)$. By assumption we have that $p(\Sigma_2 X)=\tau$ for some $X$. The question is, does there exist a $Y$ such that $\tau=p(\Sigma_1 Y)$? In other words, does there exist a $Y$ such that $\Sigma_2 X=\Sigma_1 Y$? The answer is yes; by assumption we have that $\Gamma \Sigma_1=\Sigma_2$ for some $\Gamma$. Thus, 
\begin{equation}
\Sigma_2X=\Sigma_1 Y \Rightarrow \Gamma \Sigma_1 X=\Sigma_1 Y
\end{equation}
and therefore the required $Y$ is given by 
\begin{equation}
Y=\Sigma^{-1}_1 \Gamma \Sigma_1 X
\end{equation}
Note that $Y$ does depend on both $\Sigma_1$ and $X$, as intuition would suggest.
\end{proof}

In other words, using a strictly more Blackwell-informative garbling results in a strictly larger set of feasible receiver posterior beliefs. Of course, this is obvious with trivial garblings (an identity, which would leave the feasible set unchanged from the Bayes-plausible one, and a completely uninformative garbling which would reduce the set to a single point  - just the prior), but this theorem shows that the same "nesting" is true for nontrivial Blackwell-ranked garblings. 

We illustrate (see figure 15) this observation using $\Sigma_1 = \begin{pmatrix} \frac{9}{10} & \frac{1}{100} \\ \frac{1}{10}& \frac{99}{100} \end{pmatrix}$ and $\Sigma_2 = \begin{pmatrix} \frac{2}{3} & \frac{1}{4} \\ \frac{1}{3}& \frac{3}{4} \end{pmatrix}$; it can be readily checked that $\Sigma_1 \succeq_{B} \Sigma_2$.

\begin{figure}
\includegraphics[scale=0.3]{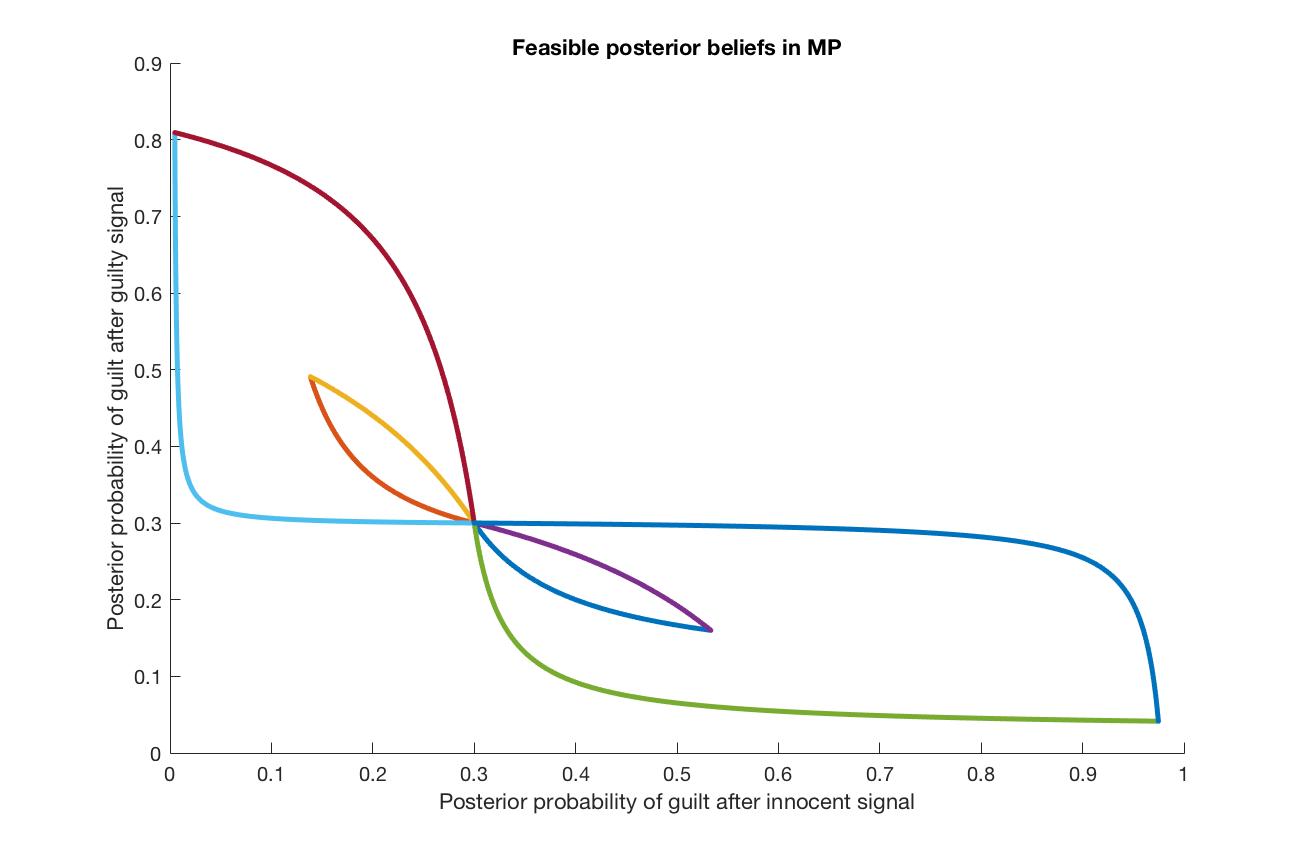}
\caption{Blackwell's Order Implies Set Inclusion for Feasible Sets.}
\end{figure}

With "filled in" convex hulls the same idea is represented in figure 16.

\begin{figure}
\includegraphics[scale=0.3]{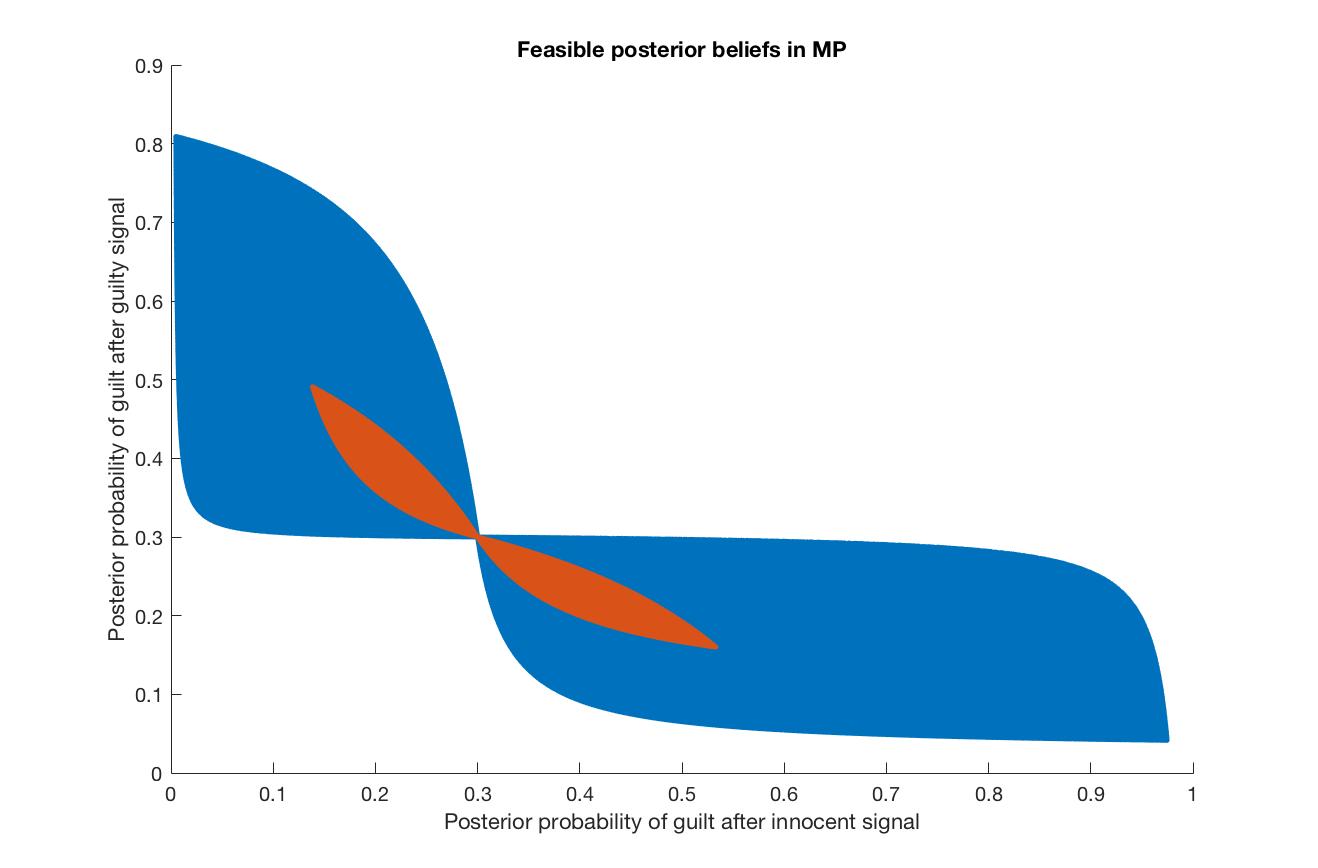}
\caption{Further Illustration of Set Inclusion.}
\end{figure}

Similarly, if $\Sigma_1$ and $\Sigma_2$ are not ranked by Blackwell's criterion, the $F$ sets are not nested. We illustrate this by an example: consider $\Sigma_1 = \begin{pmatrix} \frac{2}{3} & \frac{1}{3} \\ \frac{1}{3}& \frac{2}{3} \end{pmatrix}$ and $\Sigma_2 = \begin{pmatrix} \frac{4}{5} & \frac{1}{2} \\ \frac{1}{5}& \frac{1}{2} \end{pmatrix}$.\footnote{It can be readily checked that these matrices are not ranked.} The $F$ sets are illustrated in figure 17.

\begin{figure}
\includegraphics[scale=0.35]{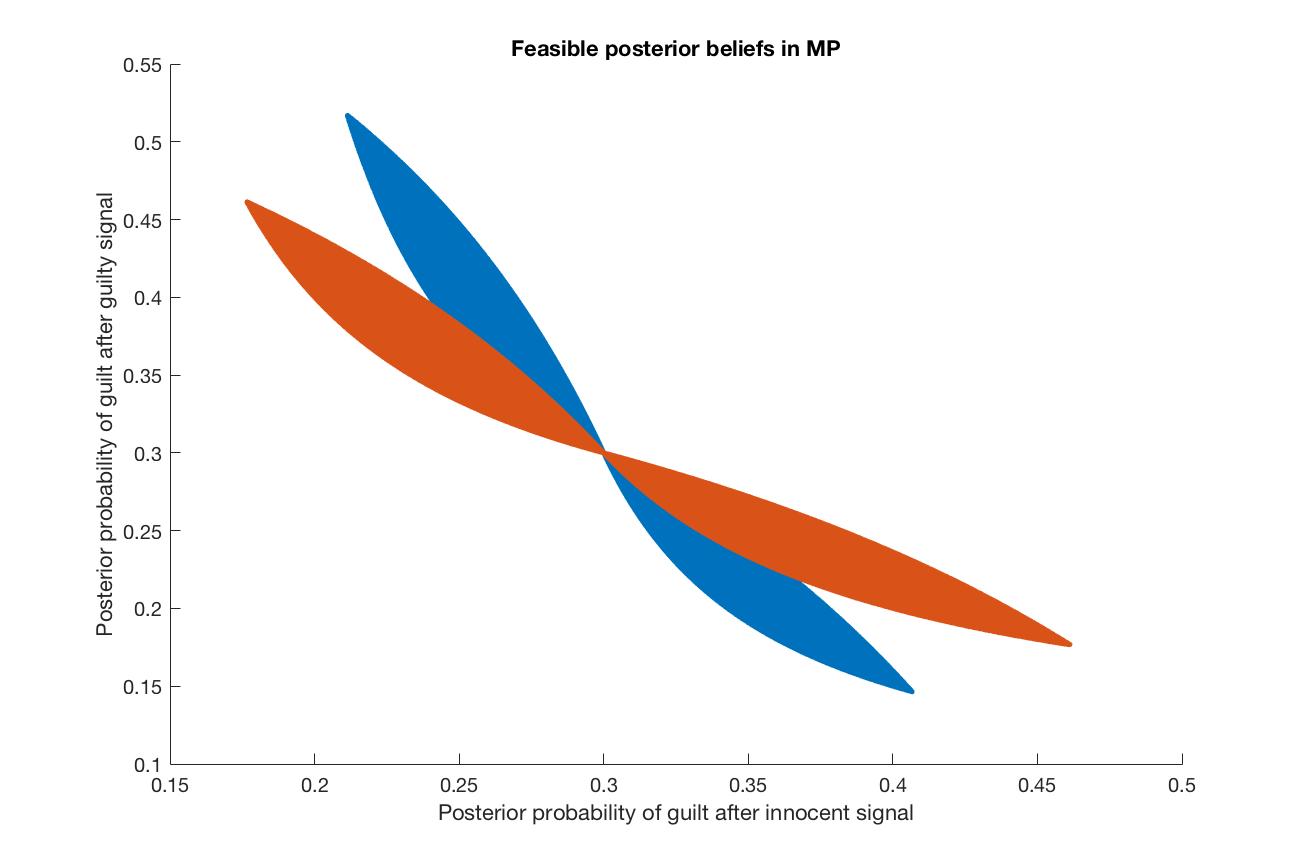}
\caption{Unranked Feasible Sets. }
\end{figure}

We now present another example to show that with two states and three signals beliefs that were not feasible with two signals, become feasible. We illustrate the set of feasible beliefs using the garbling $\Sigma =\begin{pmatrix} \frac{1}{3} & \frac{1}{9} & \frac{2}{3}\\ \frac{1}{3}& \frac{4}{9} & \frac{1}{3}\\ \frac{1}{3} & \frac{4}{9} & 0 \end{pmatrix}$. Figure 18 below demonstrates the posteriors that are feasible given this garbling. 

\begin{figure}
\includegraphics[scale=0.3]{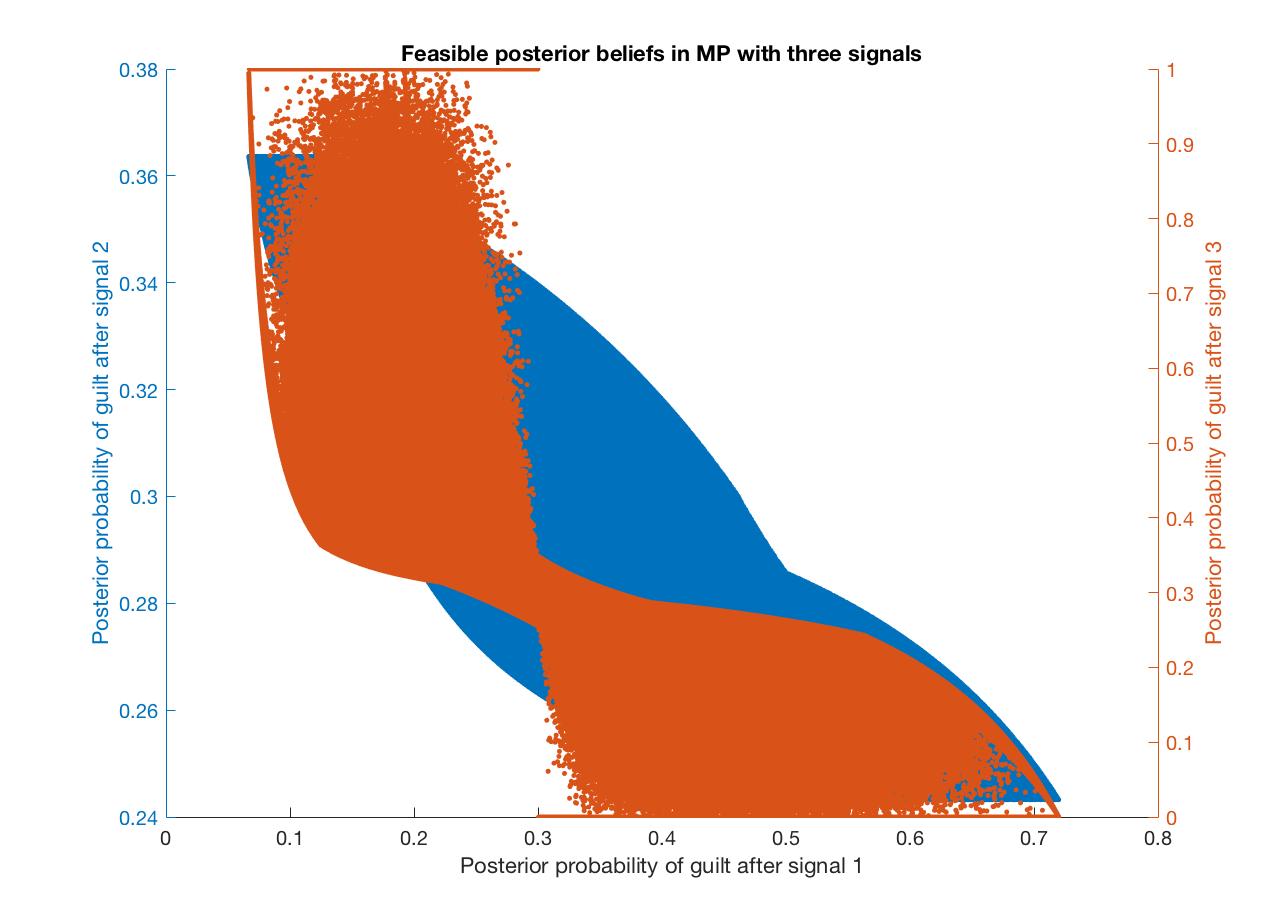}
\caption{Going Beyond the Dichotomy: Three Signals. }
\end{figure}

We have not shown all of the possible beliefs (since the sets overlap, it would be difficult to see them), but rather the outer limits of the feasible sets and some of the feasible interior beliefs. The key observation from this experiment is that with three beliefs there are beliefs that can be induced, that cannot be induced with two signals. Namely, these are beliefs below 0.3 (this can be seen by comparing the relevant figures).

\begin{subsection}{An Example Where MP Differs from BP}

We now illustrate a nontrivial example where the presence of a mediator significantly alters the baseline equilibrium. In this example the two equilibria of the mediated persuasion game are both different from the unique equilibrium of the Bayesian persuasion game. Consider a sender and a mediator with preferences illustrated in figure 19. \footnote{We relegate the discussion of the receiver's preferences and her welfare to the end.}

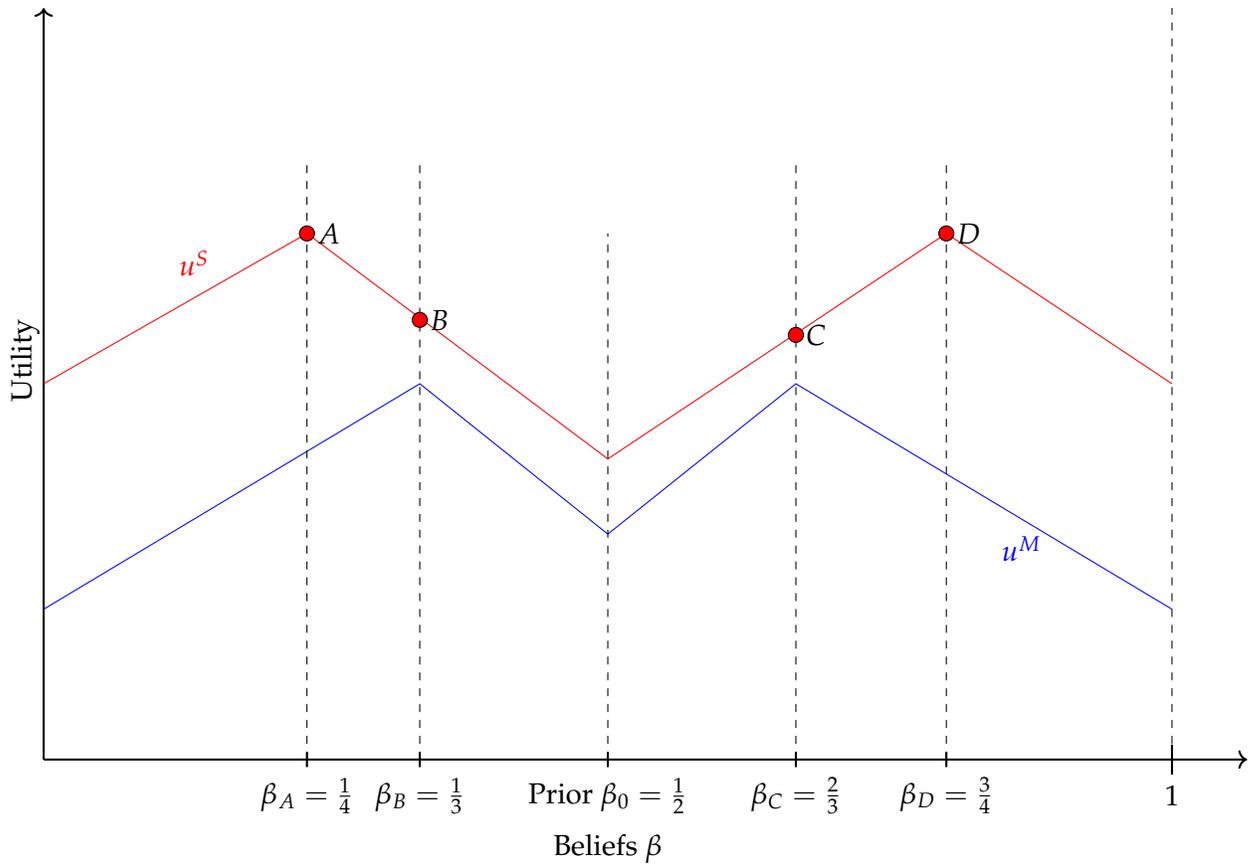
\begin{figure}[h]
\begin{tikzpicture}
\draw[thick,->] (0,0)--(0,10);
\draw[thick,->] (0,0)--(16,0);
\draw[thick] (15,-0.2)--(15,0.2);
\node [below] at (15,-0.2) {$1$};
\draw[red] (0,5)--(3.5,7)--(7.5,4)--(12,7)--(15,5);
\draw[blue] (0,2)--(5,5)--(7.5,3)--(10,5)--(15,2);
\draw[dashed] (15,0)--(15,10);
\draw[dashed] (7.5,0)--(7.5,7);
\draw[thick] (7.5,-0.1)--(7.5,0.1);
\node [below] at (7.5,-0.1) {Prior $\beta_0=\frac{1}{2}$};
\node [below] at (7.5,-0.85) {Beliefs $\beta$};
\draw[dashed] (3.5,0)--(3.5,8);
\draw[dashed] (12,0)--(12,8);
\node [left,rotate=90] at (-0.25,6) {Utility};
\draw[thick] (3.5,-0.1)--(3.5,0.1);
\draw[thick] (12,-0.1)--(12,0.1);
\draw[thick] (5,-0.1)--(5,0.1);
\draw[thick] (10,-0.1)--(10,0.1);
\draw[dashed] (5,0)--(5,8);
\draw[dashed] (10,0)--(10,8);
\node [below] at (3.5,-0.1) {$\beta_A=\frac{1}{4}$};
\node [below] at (12,-0.1) {$\beta_D=\frac{3}{4}$};
\node [below] at (5,-0.1) {$\beta_B=\frac{1}{3}$};
\node [below] at (10,-0.1) {$\beta_C=\frac{2}{3}$};
\draw[fill=red] (3.5,7) circle [radius=0.1];
\node [above, right] at (3.5,7) {$A$};
\draw[fill=red] (5,5.85) circle [radius=0.1];
\node [above, right] at (5,5.85) {$B$};
\draw[fill=red] (12,7) circle [radius=0.1];
\node [above, right] at (12,7) {$D$};
\draw[fill=red] (10,5.65) circle [radius=0.1];
\node [above, right] at (10,5.65) {$C$};
\node [above,blue] at (13,2.5) {$u^M$};
\node [above,red] at (2,6.3) {$u^S$};

\end{tikzpicture}
\caption{A Nontrivial Example: the MP Outcome is Blackwell-worse. 
The sender's utility is in red, that of the mediator is in blue}
\end{figure}

In the absence of a mediator, since her utility peaks at point $A$ and $D$, the sender would choose the posteriors $\beta_A$ and $\beta_D$ (each realizing with equal probability). However, with a mediator the situation is markedly different. In addition to the uninformative "babbling" equilibrium which always exists, there is another one in which some information is conveyed. Suppose that the mediator chooses the following signal:
$\Sigma=
\begin{pmatrix}
\frac{2}{3} & \frac{1}{3}\\
\frac{1}{3} & \frac{2}{3}
\end{pmatrix}
$. 
It can be checked (and is in fact, intuitive) that the \textit{most} informative posteriors that can be achieved given this garbling are $\beta_B$ and $\beta_C$; this is if the sender chooses a perfectly informative experiment. Any other experiment would result in a further garbling (i.e. displacement inward) of these two posteriors. Given that that sender's utility is decreasing between $B$ and $C$, it is a best response for her to indeed choose a fully revealing X, and given that this choice of $M$ is indeed optimal for the mediator, since she obtains her highest possible payoff. Thus, such and $X$ and $M$ are an equilibrium; in this equilibrium the outcome is a strict mean-preserving contraction of the outcome in the unmediated game. There are no other pure strategy equilibria in this game. 

We now turn to the question of receiver welfare; so far we have left the preferences of the receiver unspecified. It is immediate that if the receiver's preferences are the same as those of the mediator, then the receiver is strictly better off. If, on the other hand, the receiver has preferences that emphasize certainty of the state (such as the preferences of the receiver in the leading example of KG, for instance), the receiver is strictly \textit{worse} off with a mediator in this example. This simple example illustrates that the presence of the mediator has an ambiguous effect on the welfare of the receiver in general.

\end{subsection}

\begin{subsection}{Comparing Equilibrium Outcomes}
We finally come to the main part of the paper - evaluating the effect of adding a mediator to a Bayesian persuasion environment. There are a few general\footnote{I.e. those that do not depend on the exact form of the utility functions.} results that we can obtain when comparing outcomes with and without a mediator. Indeed, with a mediator, there is always a babbling equilibrium. Thus, even when there may be nontrivial persuasion/information revelation in the BP problem, there is an equilibrium without any information revelation in the MP problem, even though the preferences of the sender and the receiver are the same across the two problems. 
The second general result is that if we add a mediator whose utility is, say, globally strictly concave\footnote{Versions of this result that involve the mediator's utility being concave over some set of posteriors, or not being strictly concave are analogous and straightforward enough; we do not state them.} over the set of posteriors, not only does a babbling equilibrium exists, but it is unique. 

If sender's utility is globally strictly concave, the unique outcome in both games is no revelation. If the sender's utility is globally strictly convex, the unique outcome in both games is full revelation. If both the sender's and the mediator's utilities are linear, then any outcome can be sustained in equilibrium.

At this point, to be able to say more about outcomes, we need to start narrowing down the scope of utilities. Toward this end, suppose that the receiver's utility is convex (or at least piece-wise convex), and the sender benefits from persuasion (in the language of \hyperlink{Kamenica and Gentzkow (2011)}{Kamenica and Gentzkow (2011)}). The first result that we can state has to do with comparing two MP games where the preferences of the mediator are different; namely the set of outcomes when  $u^M=u^R$ is a subset of the set of equilibrium outcomes when $u^M=u^S$.

We now turn to comparing the informativeness of MP outcomes relative to BP outcomes and show by example that it is possible for an equilibrium of the MP game to be strictly more informative in the Blackwell sense than the equilibrium of the BP game.

Consider a sender and a mediator with preferences that are illustrated in figure 20; in this figure the sender's utility is in red and that of the mediator is in blue. The sender's utility vanishes for beliefs below 0.2, then jumps up at 0.2, jumps back down to a value of $-k$, for $k$ positive and "large", for beliefs $\beta \in (0.2, 0.955)$, except for another jump up at $0.5$ then jumps up at $0.955$, and then returns to 0. The mediator's utility is M-shaped and peaks at $0.17$ and $0.955$. The common prior is $\beta_0$; without a mediator the sender would clearly choose the posteriors $\{\beta_1^{BP},\beta_2^{BP}\}=\{0.2,0.5\}$. They are certainly Bayes-plausible, and give the sender her highest possible utility. Suppose however that the mediator chose to play the following garbling: $\Sigma = \begin{pmatrix}\frac{1}{2} & \frac{1}{100}\\ \frac{1}{2} & \frac{99}{100}  \end{pmatrix}$; the $F(\Sigma,0.3)$ set for this garbling is depicted in blue in figure 19. If the sender were to simultaneously play a fully revealing experiment, the outcome would be $\{\beta_1^{MP},\beta_2^{MP}\}=\{0.17,0.955\}$, yielding her a payoff of 0; note also that this is the most preferred outcome of the mediator. Given this garbling, the only way in which the sender can improve her payoff is by deviating to something that induces a posterior of 0.2, or 0.5. Suppose she deviates to something that results in one posterior (say, the first one) begin $\beta_1^{deviation}=0.2$. Then the second posterior must lie in the $S_1$ set also illustrated in figure 20; given this constraint, the sender would get a payoff of 1 (with some probability) when the posterior is 0.2, and the payoff of $-k$ upon the other signal realization. For $k$ large enough this will be negative for all beliefs in $S_1$, and thus this cannot be a profitable deviation. A similar logic applies to beliefs in $S_2$ and $S_3$ - inducing one belief that makes the sender better off also necessitates inducing another belief which makes her worse off, and the deviation is unprofitable. (The $S$ sets are also illustrated in the green box in figure 20; it is helpful to see them on the same figure as the utility to understand what is going on.) Obviously the mediator would not choose to deviate either, since she gets her first best outcome. Thus, the sender is "forced" to provide more information she would otherwise
\begin{equation}
\left(X,\Sigma \right)=\left(\begin{pmatrix} 1 & 0\\0 & 1\end{pmatrix}, \begin{pmatrix}\frac{1}{2} & \frac{1}{100}\\ \frac{1}{2} & \frac{99}{100}  \end{pmatrix}\right)
\end{equation}
is an equilibrium in which the outcome supported on $\{\beta_1^{MP},\beta_2^{MP}\}=\{0.17,0.955\}$ is strictly more Blackwell-informative than the BP outcome $\{\beta_1^{BP},\beta_2^{BP}\}=\{0.2,0.5\}$.

\begin{figure}[h]
\begin{tikzpicture}
\draw[thick,->] (0,0)--(0,7);
\draw[thick,->] (0,0)--(16,0);
\draw[thick,->] (0,0)--(0,-6);
\draw[thick] (15,-0.2)--(15,0.2);
\node [below] at (15,-0.2) {$1$};
\node [below] at (4.5,-0.5) {$\beta_0$};
\node [below] at (3.5,-0.5) {$\beta_1^{BP}$};
\node [below] at (7.5,-0.5) {$\beta_2^{BP}$};

\draw [ultra thick,red,dashed] (0,0.1)--(3.5,0.1);
\node [above, red] at (0.4,0.1) {$u^S$};
\draw[fill=none,red] (3.5,0.1) circle [radius=0.1];
\draw[fill=red] (3.5,5.85) circle [radius=0.1];
\draw[fill=red] (7.5,5.85) circle [radius=0.1];
\draw [ultra thick,red,dashed] (7.5,-2)--(14,-2);
\draw [ultra thick,red,dashed] (14,0.1)--(15,0.1);
\draw[fill=red] (14,0.1) circle [radius=0.1];

\draw [ultra thick,red,dashed] (3.5,-2)--(7.5,-2);
\draw [thick] (3.5,-0.1)--(3.5,0.1);
\node [below] at (3.5,-0.1) {$0.2$};
\node [below] at (4.5,-0.1) {$0.3$};

\draw[fill=none,red] (14,-2) circle [radius=0.1];

\draw [thick] (4.5,-0.1)--(4.5,0.1);
\draw [thick] (7.5,-0.1)--(7.5,0.1);
\node [below] at (7.5,-0.1) {$0.5$};
\draw[fill=none,red] (7.5,-2) circle [radius=0.1];
\draw[fill=none,red] (3.5,-2) circle [radius=0.1];

\draw [thick] (2.75,-0.1)--(2.75,0.1);
\draw [thick] (14,-0.1)--(14,0.1);
\node [below] at (2.75,-0.5) {$\beta_1^{MP}$};
\node [below] at (2.75,-0.1) {$0.17$};

\node [below] at (14,-0.5) {$\beta_2^{MP}$};
\node [below] at (14,-0.1) {$\approx 0.955$};

\draw [thick] (-0.1,-2)--(0.1,-2);
\node [left] at (-0.1,-2) {$-k$};

\draw [thick,blue] (0,1)--(2.75,3);
\draw [thick,blue] (2.75,3)--(7.5,0);
\draw [thick,blue] (7.5,0)--(14,3);
\draw [thick,blue] (14,3)--(15,0);
\node [above,blue] at (13,2.5) {$u^M$};

\node [left,rotate=90] at (-0.25,6) {Utility};

\draw [green,thick ] (2,-2.5)--(2,-7.2)--(16,-7.2)--(16,-2.5)--(2,-2.5);

\draw [gray,thick,dotted] (3,-4)--(10,-4)--(10,-3)--(3,-3)--(3,-4);
\node at (9,-3.5) {\Large $S_1$};

\draw [gray,thick,dotted] (3,-5)--(10,-5)--(10,-4.1)--(3,-4.1)--(3,-5);
\node at (9,-4.5) {\Large $S_2$};

\draw [gray,thick,dotted] (3,-6)--(15,-6)--(15,-5.1)--(3,-5.1)--(3,-6);
\node at (14.5,-5.5) {\Large $S_3$};

\draw [very thick,orange] (3.5,-3.4)--(3.5,-3.6);
\node [below] at (3.5,-3.5) {\footnotesize$0.2$};

\draw [very thick,orange] (4.75,-3.4)--(4.75,-3.6);
\node [below] at (4.75,-3.5) {\footnotesize $\approx 0.305$};

\draw [very thick,orange] (7,-3.4)--(7,-3.6);
\node [below] at (7,-3.5) {\footnotesize$\approx 0.4$};

\draw [very thick,orange] (4.75,-3.5)--(7,-3.5);

\draw [very thick,magenta] (7.5,-4.4)--(7.5,-4.6); 
\node [below] at (7.5,-4.5) {\footnotesize$0.5$};

\draw [very thick,magenta] (3.65,-4.4)--(3.65,-4.6); 
\node [below] at (3.45,-4.5) {\footnotesize$\approx 0.21$};

\draw [very thick,magenta] (4.35,-4.4)--(4.35,-4.6); 
\node [below] at (4.5,-4.5) {\footnotesize$\approx0.29$};

\draw [very thick,magenta] (4.35,-4.5)--(3.65,-4.5); 

\draw [very thick,purple] (3.5,-5.4)--(3.5,-5.6);
\node [below] at (3.5,-5.5) {\footnotesize$0.2$};
\draw [very thick,purple] (8,-5.4)--(8,-5.6); 
\node [below] at (8,-5.5) {\footnotesize$\approx 0.56$};

\draw [very thick,purple] (13.5,-5.4)--(13.5,-5.6); 
\node [below] at (13.5,-5.5) {\footnotesize $\approx 0.935$};

\draw [very thick,purple] (8,-5.5)--(13.5,-5.5);

\draw [orange,thick,dotted] (2.75,-5.7)--(2.75,7);
\draw [orange,thick,dotted] (3.5,-5.7)--(3.5,7);
\draw [orange,thick,dotted] (4.5,-5.7)--(4.5,7);
\draw [orange,thick,dotted] (7.5,-5.7)--(7.5,7);
\draw [orange,thick,dotted] (14,-5.7)--(14,7);

\end{tikzpicture}
\caption{An MP Equilibrium That Is Strictly More Informative Than the BP Equilibrium. The utility of the mediator is in blue, the utility of the sender is in red.}
\end{figure}
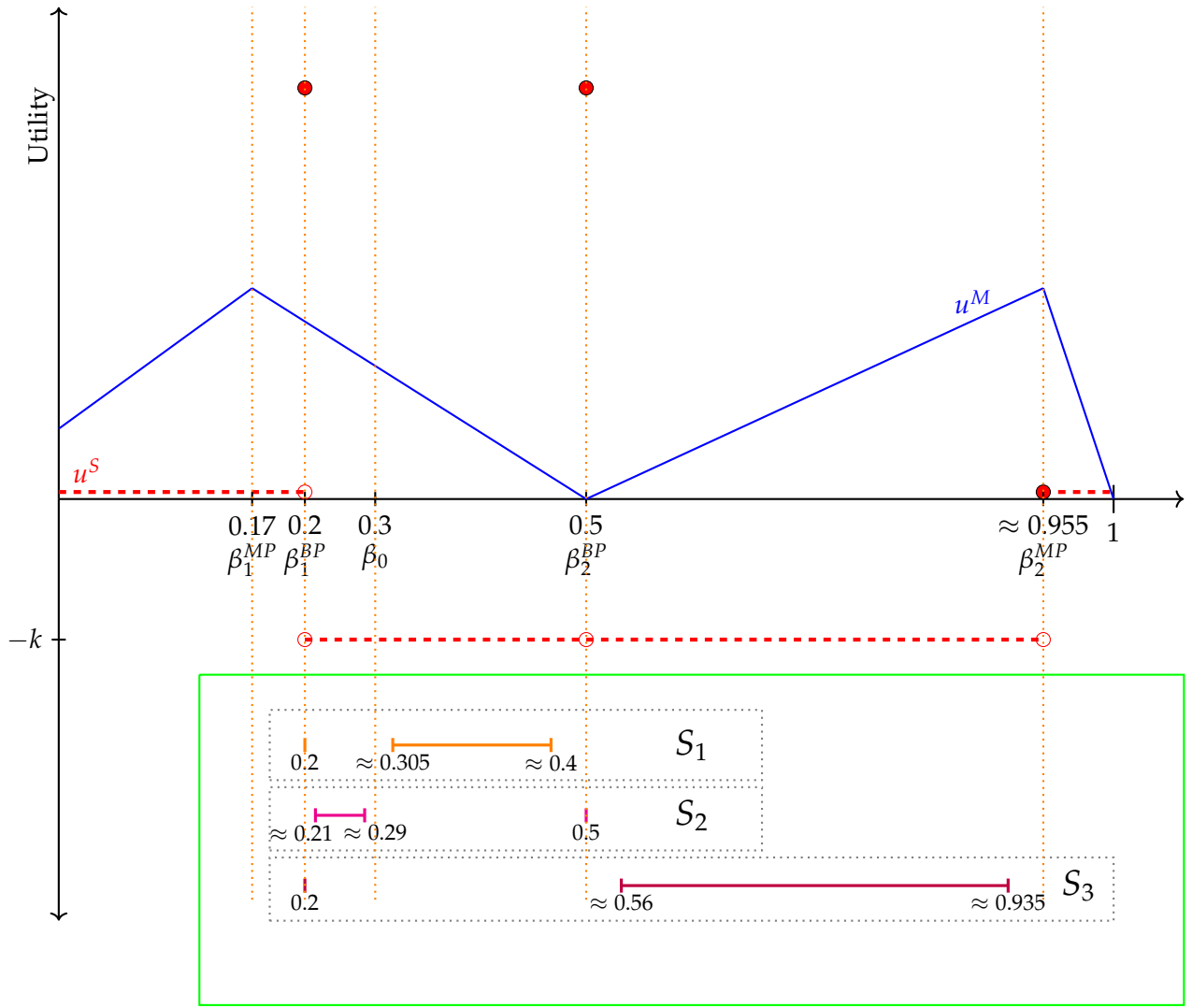

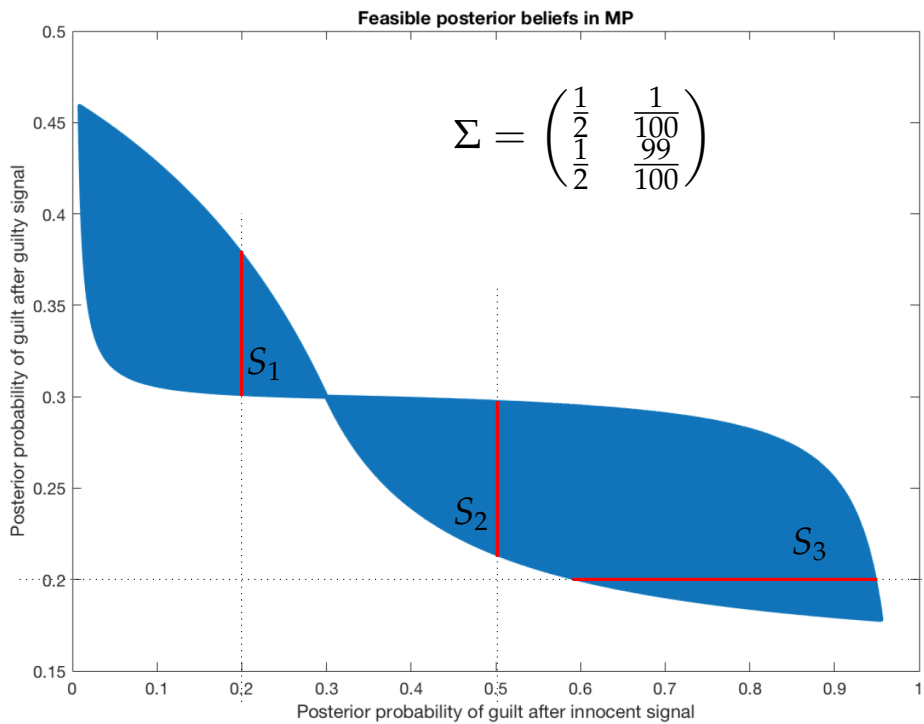
\begin{figure}
\begin{tikzpicture}
 \node[anchor=south west,inner sep=0] at (0,0) {\includegraphics[scale=0.55]{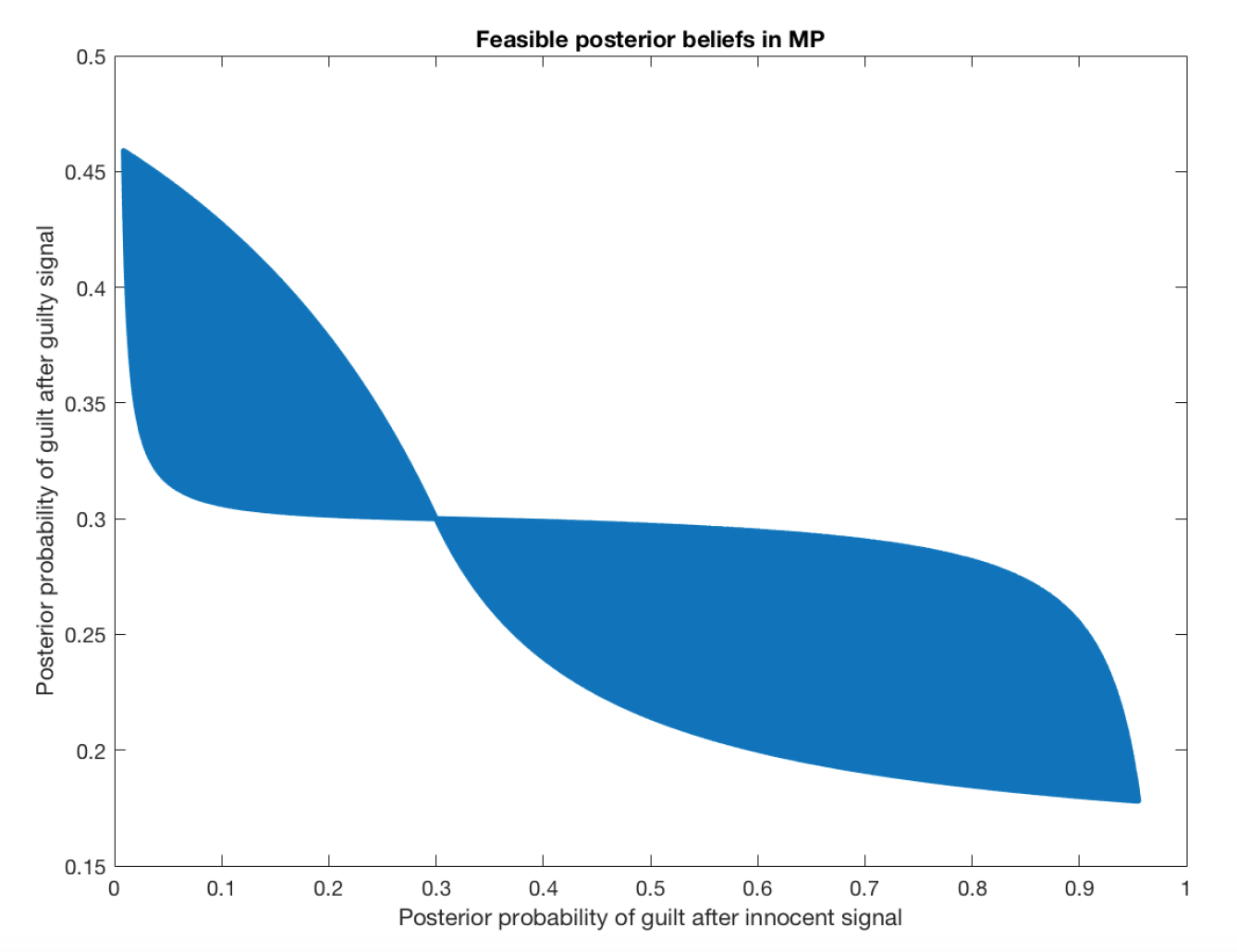}};
\node at (8,8) [scale=1.4] {$\Sigma=\begin{pmatrix}\frac{1}{2}&\frac{1}{100} \\ \frac{1}{2} &\frac{99}{100} \end{pmatrix}$};

\draw [dotted] (3.45,0.5)--(3.45,7);
\draw [red,very thick] (3.45,4.57)--(3.45,6.5);
\node at (3.75,5) {\Large $S_1$};

\draw [dotted] (0.5,2.13)--(12.5,2.13);
\draw [red,very thick] (7.85,2.13)--(11.9,2.13);
\node at (11,2.6) {\Large $S_3$};

\draw [dotted] (6.85,0.5)--(6.85,6);
\draw [red,very thick] (6.85,2.43)--(6.85,4.5);
\node at (6.5,3) {\Large $S_2$};

\end{tikzpicture}
\caption{Feasible set $F(\Sigma,0.3)$}
\end{figure}

This construction explicitly shows that there are examples where the outcome of a game where a player can only decrease the amount of information turns out to be more informative in a very strong sense. This, of course, leads to two questions - when does this happen, and what can ensure that it does not? We provide a partial answer to the second question now.\footnote{The first question is still open; we give some comments on it in the conclusion.} 

One definition would greatly simplify the exposition; we call an environment \textit{canonical} if there are only two states of the world\footnote{This is not a substantive restriction; we make this stipulation to make the intuition and the results as clear as possible.}, the receiver's utility is either strictly convex or piecewise convex, and the sender's utility can be expressed as a monotonic step function of the posterior. This subsumes the leading example of KG and covers the (natural!) situations where the receiver prefers to have as much information as possible, while the receiver has preferences that are biased\footnote{Or in the extreme case, state-independent.} toward a particular action. 

The reason for assuming monotonicity of $u^S$ is that we are focusing on the typical applications of the persuasion model where the sender has preferences that are biased in one particular direction (as is often the case in typical applications).

The sender's utility is a step function if, for example, the receiver's action set contains a finite number of elements which have some natural order or interpretation, and the sender's utility is increasing in the action. This is the case if, for example, the sender is a politician, and is trying to persuade a voter, with the voter having multiple votes to allocate among many (or one) politicians. In such a Borda rule environment each politician's utility is an increasing step function of the receiver's posterior belief. Alternatively, the sender could be a seller, and the receiver could be deciding the number of units of a good to purchase. 

This is also another reason for focusing on a step function for the sender - all other natural functional forms yield predictions that are clear enough. 
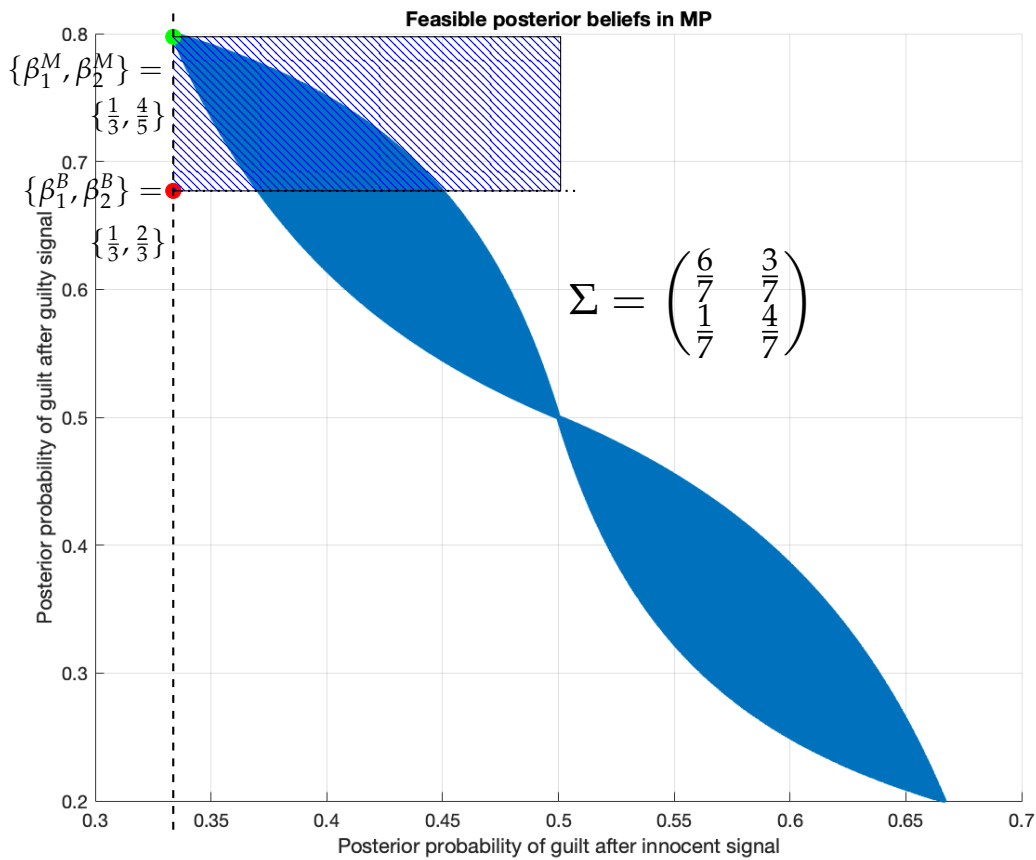
\begin{figure}[h]
\begin{tikzpicture}

\node[anchor=south west,inner sep=0] at (0,0) {\includegraphics[scale=0.35]{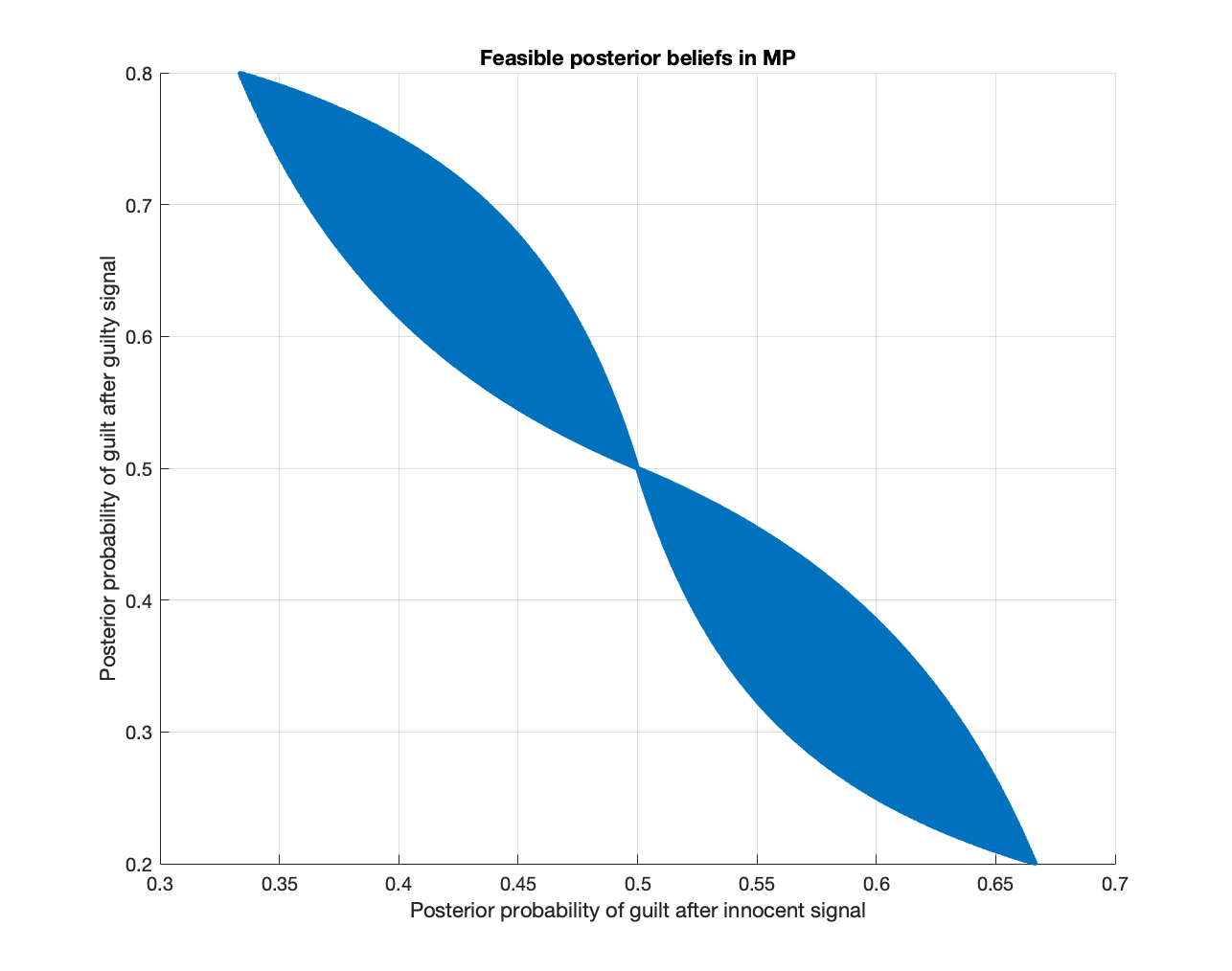}};
\node at (10,8) [scale=1.5] {$\Sigma=\begin{pmatrix}\frac{6}{7}&\frac{3}{7} \\ \frac{1}{7} &\frac{4}{7} \end{pmatrix}$};

\node [left] at (3.1,9.5) {$\{\beta_1^B,\beta_2^B\}=$};
\node [left] at (3.15,8.8) {$\{\frac{1}{3},\frac{2}{3}\}$};
\draw[fill=red,red] (3.1,9.5) circle [radius=0.1];

\draw[fill=green,green] (3.1,11.55) circle [radius=0.1];
\node [below,left] at (3.1,11.1) {$\{\beta_1^M,\beta_2^M\}=$};
\node [left] at (3.15,10.5) {$\{\frac{1}{3},\frac{4}{5}\}$};

\draw [thick,dashed] (3.1,1)--(3.1,11.9);

\draw [thick, dotted] (3.1,9.5)--(8.5,9.5);

\draw[pattern=north west lines, pattern color=blue] (3.1,9.5) --(8.25,9.5) -- (8.25,11.55) -- (3.1,11.55);

\end{tikzpicture}
\caption{An Informative MP Equilibrium in a Canonical Environment.}
\end{figure}
\begin{theorem}
Suppose that the environment is canonical, $|A|=2$, and the receiver's pure strategy is onto. Then the outcome of the MP game cannot be strictly Blackwell-more informative than the outcome of the BP game.
\end{theorem}
The reason for phrasing the theorem thus is twofold. First of all, we are looking to highlight the fact that MP result in \textit{strictly} more information being revealed - in other words, there is no reason to prefer to have a mediator under these assumptions. Secondly, with linear utilities for both the sender and the mediator we can trivially support any outcome - since both players are indifferent over everything - so it is very easy to come up with examples where the outcomes are exactly identical. 
\begin{proof}
Suppose first that $A=\{a_1,a_2\}$, so that there are two actions the receiver can take; since the receiver's strategy is onto she also takes both actions for some beliefs. Suppose (wlog) that she takes action $a_1$ for posterior beliefs $\beta(\sigma)\in C_1 \triangleq [0,c_1)$ and action $a_2$ for $\beta(\sigma)\in C_2 \triangleq [c_1,1]$.\footnote{We have implicitly assumed that when she is indifferent she takes the sender's most preferred action.} Thus the sender's utility can be written as $u^S(\beta)=k_1 \mathbb{1}_{\{\beta(\sigma)\in C_1\}}+k_2\mathbb{1}_{\{\beta(\sigma)\in C_2\}}$ for some $k_1,k_2$ with $k_1<k_2$. Suppose first that the prior  $\beta_0 \in C_2$; for future reference, we refer to this case as \textit{base case 2}. Then the BP game either has a unique equilibrium (if $\beta_0=c_1$) or any outcome $\{\beta_1^B,\beta_2^B\}$ with $c_1\leq \beta_1^B\leq\beta_0\leq\beta_2^B\leq1$ can be supported. Thus, the only way for an MP outcome to be more informative than the BP outcome is to have one of the beliefs (say, $\beta_1^M$) be below $c_1$. But then the sender's utility would be lower than $k_2$ which is what is in any BP equilibrium if $\beta_0 \in A_2$, since it would have to be some combination of $k_1$ and $k_2$ and $k_1<k_2$ by monotonicity. Therefore there always exists a profitable deviation for the sender - choose an uninformative experiment, and bring the utility back up to $k_2$; this is always feasible and is strictly better so in this case an MP outcome cannot be strictly more informative than the BP outcome. 

Suppose now that $\beta_0 \in C_1\setminus\{0\}$; for future reference, we refer to this case as \textit{base case 1}. Note that in this case there is a unique equilibrium in the BP game:$\{\beta_1^B,\beta_2^B\}=\{0,c_1\}$.\footnote{Note also that in this case after one of the signal realizations the receiver will take a worst action for the sender, and thus, by a theorem from KG, she will be certain of her action in that case.} The only way for an MP outcome to be more informative is to have $\beta_2^M>c_1$.\footnote{Of course, we also have $\beta_1^M=0$.}; denote by $\{X^*,\Sigma^*\}$ the choices leading to such an outcome. But then we would have $\{0,c_1\}\notin F(\Sigma^*,\beta_0)$ and yet $\{0,\beta_2^M\}\in F(\Sigma^*,\beta_0)$. Direct computation (which we omit) shows that this is impossible. Finally, if $\beta_0=0$, i.e. the prior vanishes to begin with, then the only Bayes-plausible posteriors are uninformative in both the BP and the MP cases, and thus, one cannot be strictly more informative than the other. This finishes the proof for the case where $|A|=2$.
\end{proof}

We now show by example that even in this simple and canonical setting the MP outcome can be strictly more informative than the BP outcome. Suppose that the common prior is $\beta_0=\frac{1}{2}$, that $A=\{a_1,a_2,a_3\}$, and that the optimal strategy of the receiver is to take action $a_1$ for $\beta(\sigma)\in [0,\frac{1}{3})$, take action $a_2$ for $\beta(\sigma)\in [\frac{1}{3},\frac{2}{3})$, and take action $a_3$ for $\beta(\sigma)\in [\frac{2}{3},1]$. By monotonicity of the sender's utility it follows that she would induce posterior beliefs $\tau^{BP}=\begin{cases} \frac{1}{3} \text{ with probability } \frac{1}{2}\\ \frac{2}{3}\text{ with probability } \frac{1}{2}\end{cases}$, using the experiment $X^{BP}=\begin{pmatrix}\frac{2}{3}&&\frac{1}{3}\\ \frac{1}{3}&&\frac{2}{3} \end{pmatrix}$.

We claim, however, that with the appropriate preferences for the mediator, the pair $\{X^*,\Sigma^*\}$ can be an equilibrium, where 
\begin{equation}
X^*=\begin{pmatrix}1&&0\\0 && 1 \end{pmatrix}, \Sigma^*=\begin{pmatrix}\frac{6}{7}&&\frac{3}{7}\\ \frac{1}{7} && \frac{4}{7} \end{pmatrix}
\end{equation}
yielding the following distribution of posterior beliefs: $\tau^{MP}=\begin{cases} \frac{1}{3} \text{ with probability } \frac{9}{14}\\ \frac{4}{5}\text{ with probability } \frac{5}{14}\end{cases}$. It can be checked (and is in fact, graphically apparent from figure 22) that $\tau^{MP}$ is a mean-preserving spread of $\tau^{BP}$, so that the MP outcome is Blackwell more informative than the BP outcome. To show that this is an equilibrium, we check that nobody can profitably deviate. Suppose first that the mediator is choosing $\Sigma^*$; $F(\Sigma^*,\frac{1}{2})$ is depicted in figure 22. In the shaded region the sender's utility is increasing in the southwestern direction, so for $k_3$ high enough the best that she could do is induce the green point (using a fully revealing information structure). And given that the sender is choosing full revelation, it is easy to construct mediator preferences that would result in $\Sigma^*$ being the optimal choice. 




We say that \textit{the receiver benefits from mediation} if the receiver's utility in the mediated persuasion game is strictly greater than her utility in the Bayesian persuasion game, where of course we compare games where the preferences of the sender and the receiver do not change between games. 

We now show a surprising result; when the preferences of the mediator are perfectly aligned with those of the sender (or alternatively, the receiver simply \textit{is} the mediator), the receiver cannot benefit from mediation. The import of this finding is that the receiver benefits from mediation when the mediator's preferences are such that the mediator prefers more information revelation, in the sense of Blackwell, than the sender, but less information revelation than the receiver (which in a canonical environment, is of course, full revelation). 
\begin{theorem}
Suppose that the environment is canonical, and furthermore that $u^M=u^R$; suppose to avoid trivialities that the sender also benefits from persuasion. We now show that if $\tau^*$ with support equal to $\{\beta_1^*,\beta_2^*\}$ is an equilibrium outcome, then we cannot have $\tau^* \succ \tau^{BP}$.
\end{theorem}
\begin{proof}
Suppose, towards a contradiction, that $\tau^* \succ \tau^{BP}$, and denote by $(X^*,\Sigma^*)$ the information structures chosen. By supposition, $\Sigma \neq I_2$, since otherwise we would have $\{\beta_1^*,\beta_2^*\}=\{\beta_1^{BP},\beta_2^{BP}\}$; so the mediator must be choosing some nontrivial garbling. Note that this also implies that $X^* \succ X^{BP}$. Thus, since the mediator's utility is convex, deviating to $\Sigma'=I_2$ yields a strictly higher utility, and thus, $(X^*,\Sigma^*)$ could not have been an equilibrium.
\end{proof}

In other words, if $u^M=u^R$, the receiver cannot strictly benefit since if that were the case, the mediator would have a profitable deviation. And that profitable deviation - no garbling at all - cannot be part of an equilibrium where $\tau^* \succ \tau^{BP}$ since then the sender would have a profitable deviation, bringing beliefs back to the BP outcome, which must now be feasible, since the mediator is not garbling the information at all, and thus all Bayes-plausible beliefs are feasible. Thus, if the sender were the mediator, there can exist a garbling that makes her better off, but that cannot be an equilibrium, since there is also always a better one still until we get to no garbling at all, which again cannot be part of an equilibrium. In fact, remarkably enough, it can be shown that the outcome when $u^M=u^R$ is either uninformative (i.e. $\beta_1^*=\beta_2^*=\beta_0$), or it coincides with the BP outcome\footnote{The argument is simple and goes like this - we have shown that we cannot have $\tau^* \succ \tau^{BP}$ strictly; we also cannot have $\tau^{BP} \succ \tau^{*}$ since then the mediator would have a profitable deviation. Finally, if the posteriors were unranked, the mediator would again move one of the beliefs outward until it at least coincides with the BP belief. This exhausts the possibilities and proves the result.}. In other words, in the dichotomy setting, the nontrivial outcome when $u^M=u^S$ or when $u^M=u^R$ is exactly the same. 

We have established two important points: 1) the receiver can be better off in a very strong sense with a mediator, even though the mediator can only destroy information, and 2) for this to happen the preferences of the mediator cannot be the same as the preferences of the receiver (or the sender, for that matter). It must be the case that the mediator prefers more information revelation than the sender, but not perfect revelation (which of course is the preferred outcome of the receiver by assumption). 
\end{subsection}

\begin{subsection}{Interpretation of the Rank of a Garbling Matrix}
We now turn to a discussion of one of the key conditions established above - the necessity for $\Sigma$ to be of full rank. This is a fairly straightforward question, yet it has never come up in the literature - what is the economic interpretation of the rank of a garbling matrix?

For simplicity suppose that the matrix is square, so that full rank guarantees invertibility. We first start with a discussion of what is means for a garbling matrix to \textit{not} be invertible. By definition of rank, the column rank and the row rank of a matrix are always identical; recall also the convention that the columns of a garbling matrix represent signal realizations in each state of the world. If a matrix is not invertible, it means that there is at least one columns (a profile of signals in a given state) that is a linear combination of the other columns. In other words, one can \textit{replicate the distribution of signals in a state without knowing anything about the state}. This is literally the definition of a Blackwell garbling. 

The corresponding (row) point of view offers the same insight. If a garbling matrix is not invertible, then the distribution of a particular signal in all possible states is a linear combination of the distributions of the signals in the other states, and hence, one can replicate the distribution of a signal. In other words, a singular garbling contains within itself a sort of Blackwell garbling. Whether or not this internal garbling can be "undone", perhaps by constructing a new one, remains an open question\footnote{For example, given a garbling suppose that the receiver constructs another garbling from that, one that has full rank. What are the properties of this artificial garbling relative to the original one?}

This discussion sheds some light on the invertibility condition. The fact that the garblings used in the discussion of the feasible sets were all invertible means that they carry "as much information as possible", given their dimensional constraints. 

Finally, suppose that the garbling is not square, i.e. $\Sigma$ is a $m$-by-$n$ matrix with $m$ signals, $n$ states and $m \geq n$.\footnote{Recall that the assumption that there are at least as many signals as states is made to avoid some trivialities which arise when the signal space is not "rich enough".}  $\Sigma$ being full rank means that the rank is equal to $n$ (the most it can be), the number of states, which in turn implies that there always exists a \textit{left} inverse. Observe that all of the inverses discussed so far were always used in left-multiplying the relevant matrices, so for non-square garblings the logic and algebra of being full rank is the same as the logic of invertibility for square matrices.

\end{subsection}

\end{subsection}

\end{section}






\end{section}

\begin{section}{Concluding Remarks: Towards a Characterization of Equilibrium Outcomes}

A major, and surprising, finding of this paper has been that the receiver benefits from mediation when the mediator's preferences are more dispersed than those of the sender. Interestingly, if the mediator's preferences are the same as the receiver (i.e. also convex), then the receiver does \textit{not} benefit from mediation. An implication of this finding, is, of course, that if one were to decide whether or not to include a mediator in a setting such as ours, and one were inclined to choose outcomes that are preferred by the receiver, then one would choose a mediator only if such a mediator's preferences were sufficiently \textit{different} from the receiver's. 

We now comment informally on what a characterization equilibrium outcomes might look like. The first observation that one can make is that if a belief is part of an equilibrium outcome, it must be the case that at that belief the utility of a player coincides with the concavification of the utility; Matthew Gentzkow and Emir Kamenica refer to such beliefs as "coincident" throughout their work on this topic. This narrows down the set of possible beliefs considerably. 

The second important observation is that if there is to be any information revelation, the utilities of the players have to be convex over a set that includes the prior. Let us call a set of posterior beliefs of a player a \textit{convex basin} if is the set over which the utility is weakly convex. Putting these two observations together, we obtain a hint of what characterization of equilibrium outcomes might look like - it must be the intersection of convex basins\footnote{Or at least a union of convex basins, to account for the possibility of an inverse-M-shaped utility function.} of players' utilities, and a belief can be part of an equilibrium outcome only if is coincident over this intersection of convex basins. While a formal statement is not available at the moment, it seems to be within our grasp.

The future of this project will address sequential versions of this model - namely, what happens when we allow the mediator to move before or after the sender, and perhaps, condition the mediator's choice of action upon the experiment chosen by the sender, while still maintaining the assumption of fixed signal realization spaces. In other words, we will work on understanding sequential persuasion where players can also destroy information, not only add it, as has been the case up to now. 

\end{section}

\end{section}

\end{document}